\documentclass[]{aa}
\usepackage{linenoaa}
\providecommand{\dodoi}[1]{doi:~\href{http://doi.org/#1}{\nolinkurl{#1}}}
\providecommand{\doarXiv}[1]{arXiv:~\href{https://arxiv.org/abs/#1}{\nolinkurl{#1}}}
\nolinenumbers
\DeclareRobustCommand{\VAN}[3]{#2}
\let\VANthebibliography\thebibliography
\def\thebibliography{\DeclareRobustCommand{\VAN}[3]{##3}\VANthebibliography}

\usepackage{academicons}
\usepackage{xcolor}
\usepackage{orcidlink}
\usepackage{tablefootnote}
\usepackage{subfig}
\usepackage{hyperref}
\usepackage{natbib}
\bibpunct{(}{)}{;}{a}{}{,} % to follow the A&A style

\def\arcmin{\hbox{$^\prime$}}
\def\arcsec{\hbox{$^{\prime\prime}$}}
\def\approxlt{\ifmmode \rlap{$<$}{}_{{}_{{}_{\textstyle\sim}}} \else%
$\rlap{$<$}{}_{{}_{{}_{\textstyle\sim}}}$\fi}
\def\chan{{\it Chandra}}
\def\xmm{XMM-{\it Newton}}
\def\swf{{\it Swift}}

\def\degr{\hbox{$^\circ$}}
\def\arcmin{\hbox{$^\prime$}}
\def\arcsec{\hbox{$^{\prime\prime}$}}

\usepackage{lineno}
\linenumbers

\usepackage{hyperref}
\usepackage{academicons}
\usepackage{xcolor}
\usepackage{orcidlink}
\usepackage{tablefootnote}
\usepackage{longtable}

\begin{document}
\titlerunning{IMBH J1231 with a varying accretion rate}
\authorrunning{Cao et al.}

\title{The intermediate--mass black hole 2XMM J123103.2+110648: a varying disc accretion rate during possible X--ray quasi--periodic eruptions?}

\author{Zheng Cao\orcidlink{0000-0002-0588-6555}
\inst{1,2}\fnmsep\thanks{z.cao@sron.nl},
Peter G.~Jonker\orcidlink{0000-0001-5679-0695}\inst{2, 1},
Sixiang Wen\orcidlink{0000-0002-0934-2686}\inst{3},
Nicholas C.~Stone\orcidlink{0000-0002-4337-9458}\inst{4,5},
\and
Ann I.~Zabludoff\orcidlink{0000-0001-6047-8469}\inst{6}}

\institute{SRON, Netherlands Institute for Space Research, Niels Bohrweg 4, 2333 CA, Leiden, The Netherlands
\and
Department of Astrophysics/IMAPP, Radboud University, P.O.~Box 9010, 6500 GL, Nijmegen, The Netherlands
\and
National Astronomical Observatories, Chinese Academy of Sciences, 20A Datun Road, Beijing 100101, China
\and
Department of Astronomy, University of Wisconsin, Madison, WI 53706, USA
\and
Racah Institute of Physics, The Hebrew University, Jerusalem, 91904, Israel
\and
The University of Arizona, 933 N. Cherry Ave., Tucson, AZ 85721, USA
}

 \date{Received XXX; accepted XXX}

\abstract{
We fit the evolving X-ray spectra of the variable and fading source 2XMM~J123103.2+110648 (J1231), which is an intermediate--mass black hole (IMBH) candidate. Recent X--ray timing studies have proposed that J1231’s quasi-periodic oscillation (QPO) observed at the peak of its X-ray lightcurve is a variant of the quasi--periodic eruptions (QPEs) observed in other sources. Here, we fit X--ray spectra from \xmm{}, \swf{}, and \chan{} using a slim disc model for the black hole’s accretion disc, obtaining a best-fit black hole mass of ($6\pm3)\times10^{4}$~$M_\odot$ and spin of $>0.6$ at 2$\sigma$ confidence. This mass is consistent with past estimates, supporting the IMBH interpretation, and the spin measurement is new. Yet the nature of J1231 remains uncertain: its long-term variability (decade-long continuum evolution) could signal a tidal disruption event or active galactic nuclear variability. We find that the spectral evolution within the first three years after the source's detection can be well explained by either a varying disc accretion rate $\dot m$ or a varying disc inclination $\theta$. Meanwhile, we find that during the short-term variability (the QPO with a $\sim3.8$~hr period), each oscillation does not show the "hard--rise--soft--decay" typical of QPEs. We fit the average spectrum at the QPO lightcurve maxima and the average spectrum at its minima, finding that the spectral difference is well explained by $\dot m$ decreasing from peaks to valleys if $\theta<30^{\circ}$ and constant between all data epochs. This result suggests that the short--term QPO behaviour might also be driven by a varying disc $\dot m$.}

\maketitle

%%%%%%%%%%%%%%%%%%%%%%%%%%%%%%%%%%%%%%%%%%%%%%%%%%

%%%%%%%%%%%%%%%%% BODY OF PAPER %%%%%%%%%%%%%%%%%%
\nolinenumbers
\section{Introduction}
\label{sec:intro}

Intermediate--mass
($10^{2}$--$10^{5}$~M$_\odot$)
black holes (IMBHs) are believed to play a vital role in the formation history of supermassive black holes (SMBHs; $\gtrsim10^{6}$~$M_\odot$; e.g., \citealt{volonteri2010formation,kormendy2013coevolution,natarajan2014seeds,shankar2016selection,pacucci2018glimmering,banados2018800}). Measuring the mass and spin distributions of IMBHs can help us understand the collective formation and evolutionary history of IMBHs and SMBHs \citep[e.g.,][]{greene2020intermediate,inayoshi2020assembly}. 2XMM~J123103.2+110648 (J1231; redshift $z=0.11871$; \citealt{ho2012low}) is an accreting IMBH candidate. It was serendipitously discovered in archival \xmm{} X--ray data \citep{terashima2012candidate,lin2013classification}. 
The source X-ray flux decayed by $\approx1$ order of magnitude over the time period 2006 to 2016 \citep{lin2017large}. Optical data indicate that the source could be an IMBH; the BH mass is derived using the observed line width extrapolating the empirical relation between the BH mass and the velocity dispersion of optical lines from the galaxy ($\sim1\times10^5$~$M_{\odot}$; \citealt{ho2012low}).

The origin of the changes in J1231's X-ray emission is unclear. It has been proposed that J1231 is a tidal disruption event (TDE; \citealt{lin20133,lin2017large}), i.e., a star that has approached and then been tidally disrupted by the black hole, leading to the formation of an accretion disc \citep[e.g.,][]{hills1975possible,rees1988tidal}. 
The behaviour of J1231's X-ray emission supports this picture:
both 
staying at high X--ray luminosity for years ($>10^{41}$erg$/$s between 0.3--10~keV)
\citep[e.g.,][]{rees1988tidal,maksym2014rbs,lin2017large,wen2020continuum} and the very soft (most photons are $\lesssim2$~keV) X--ray spectra are in line with typical TDE lightcurves and spectra, which are dominated by disc emission \citep[e.g.,][]{ulmer1999flares,lodato2011multiband,lin20133,lin2017large,guolo2024systematic}.

\begin{table*}
\centering
\scriptsize
\caption[]{Journal listing properties of the archival observations of J1231 used in this work.}
\begin{tabular}{cccccc}
\hline
Satellite  & ObsID (Label)    & Date   & Exposure (ks) & Energy range (keV) &  Est.~\# Source counts \\
\hline
\xmm{} & 0145800101 (X1) & 2003/07/13 & 36.5 (pn) / 45.3 (MOS) & 0.3--1.0 & 1026 \\
           & 0306630101 (X2) & 2005/12/13 & 52.9 (pn) / 65.0 (MOS) & 0.3--2.0 & 3094\\
           & 0306630201 (X3) & 2005/12/17 & 80.6 (pn) / 90.8 (MOS) & 0.3--2.0 & 3459\\
\hline
\swf{} & 00032732001--00032732011 (S1) & 2013/03/08--2014/07/27 & 51.2 (XRT) & 0.3--1.0 & 17 \\
\hline
\chan{} & 17129 (C1) & 2016/02/10 & 39.5 (ACIS) & 0.3--7.0 & 9 \\
\hline
\end{tabular}
\tablefoot{Labels of observing epochs are given in brackets following the observation ID. \swf{} obtained 11 exposures between 2013 March and 2014 July. In our analysis, we average all those data and treat them as a single epoch (Epoch S1). For the \xmm{} epochs, we give the exposure after filtering out periods of enhanced background radiation. This is done separately for each of the two instruments used, as specified in between brackets following the exposure time. For each spectrum, the energy band we use in our spectral analysis is also given, as we discard the data bins where the background count rate is larger than the source count rate. In the last column we estimate the net source counts in the given energy bands (the \xmm{} pn and MOS counts are added).}
\label{tb:da}
\end{table*}

It is also possible that J1231's long--term changing emission over the decade arises from active galactic nucleus (AGN) variability. Optical spectra from the host galaxy taken in 2012 indicate the presence of a low--luminosity ($g$--band magnitude -17.9~mag), type--2 AGN \citep{ho2012low,lin2017large}, while a TDE--associated origin of the narrow optical lines is not excluded. As stressed by \citet{lin2017large}, the observed long--term variability may be due to an AGN disc instability (as proposed for NGC~3599 and IC~3599; e.g., \citealt{saxton2015soft,grupe2015ic,inkenhaag2021host}).

Adding to the intrigue is J1231's X--ray quasi--periodic short-term variability, over a timescale of hours, that is observed 
during two of the three \xmm{} epochs. \xmm{} observed J1231 on 2003-07-13, 2005-12-13, and 2005-12-17\footnote{ObsID: 0145800101, 0306630101, and 0306630201, respectively.}. In this paper, we label these three epochs as X1, X2, and X3, respectively. \citet{lin20133} find X--ray quasi--periodic oscillations (QPOs) with a $\sim3.8$~hr period at X2 and X3, but at X1 the QPO is not turned on yet. Observations $\approx10$ years later by \chan{} and \swf{} do not show such short--term variability, suggesting that the QPO has either turned off or that it has become undetectable due to the diminished source flux (see Table~\ref{tb:da} for the estimated flux at each observation epoch). 
It is possible that the J1231's QPO is related to the low--frequency QPOs (LFQPOs; e.g., \citealt{remillard2006x}) detected in X--ray binaries of stellar--mass BHs, but scaled--up to the IMBH mass regime \citep{lin20133}. LFQPOs in X-ray binaries are sometimes attributed to Lense-Thirring precession of a misaligned accretion disk, which would arise naturally in a TDE \citep{stone2012observing}.

Recently, a new form of X--ray variability --- quasi--periodic eruptions (QPEs)--- has been discovered in several TDEs and AGNs \citep[e.g,][]{miniutti2019nine,giustini2020x,arcodia2021x,chakraborty2021possible,evans2023monthly,quintin2023tormund,nicholl2024quasi,arcodia2024more,guolo2024x}. While the physical origin of QPEs is actively debated, they are observed as rapid X--ray flares ($\sim$ks) separated by quiescent baselines, differentiated from the gentler, quasi--sinusoidal modulations of standard QPOs. Even though J1231's QPOs do not show a clear distinction between the flares and quiescence, 
the variability timescales are similar to QPEs, so 
it has been proposed that J1231 is a QPE variant \citep[e.g.,][]{webbe2023variability,king2023qpe}. Investigating the physical origin of J1231's short--term variability could shed light on both mechanisms responsible for X--ray variability, and the possible link between J1231 and QPE sources.

In this paper, we fit a decade of evolving J1231 X--ray spectra to constrain the black hole mass and spin with our slim disc model of the accretion disc. As the X--ray timing analysis for X2 and X3 has been performed in previous work \citep{lin20133,webbe2023variability}, we focus on spectral analysis here. This paper is structured as follows: In Section 2, we describe the data and data reduction methods. In Section 3, we present the results from our analysis. In Section 4, we discuss the implications of our results. In Section 5, we present our conclusions.

\section{Observations and methods}
\label{sc:data}

J1231 has been observed in X--rays by several satellites since the outburst start in 2003. In our analysis, we use all the archival X--ray spectroscopic data of J1231 available by the end of 2024. A journal of the data used is given in Table~\ref{tb:da}. The details of the data reduction process are presented in the Appendix.

We carry out spectral analysis using the {\sc XSPEC} package \citep{arnaud1996xspec} version 12.14.0. For consistency, we create a logarithmic energy array of 1000 bins from 0.1 to 100.0~keV for model calculations in all analysis ({\sc energies} command in {\sc XSPEC}). When fitting models to data, we evaluate the goodness--of--fit using Poisson statistics (\citealt{cash1979parameter}; {\sc C-STAT} in {\sc XSPEC}), due to the low photon counts ($<$100) in some of the spectra. We calculate the expected value of C-statistic, $C_e$, and its root--mean--square (RMS), based on the polynomial expressions given by \citet{kaastra2017use}. We re--bin every background and source$+$background spectrum by the optimal--binning algorithm (\citealt{kaastra2016optimal}; using the FTOOL {\sc ftgrouppha}), while requiring the spectra to have a minimum of 1 count per bin (with parameter {\sc grouptype} set to {\sc optmin} in {\sc ftgrouppha}). For each spectrum, we discard the data bins where the background count rate is higher than the source count rate. The energy range in each spectrum that remains after this filtering is listed in Table~\ref{tb:da} for each observation. Unless mentioned otherwise, we quote all parameter errors at the 1$\sigma$ (68\%) confidence level, corresponding to $\Delta$C-stat = 1.0 and $\Delta$C-stat=2.3 for single-- and two--parameter error estimates, respectively.

We first fit the background spectrum.
The background fit function is phenomenological, and it consists of up to two Gaussian components and one to three power--law components (depending on the instrument). The full--width at half--maximum (FWHM) of each background Gaussian component is fixed to $\sigma_{\rm Gauss}=0.001$~keV, this is less than the spectral resolution of all instruments considered in this paper. This phenomenological background model accounts for both a background continuum and possible fluorescence lines \citep[e.g.,][]{katayama2004properties,pagani2007characterization,harrison2010nuclear}. 
Next, we add the best--fit background model to the fit function describing the source$+$background spectrum. The background model parameters are kept fixed at the best--fit values determined from the fit to the background--only spectrum. In this paper, when studying the source$+$background spectra, we refer to the part of the fit function that describes the source as \textit{fit function}.

In the spectral analysis, we include the Galactic absorption in the fit function through the {\sc XSPEC} model \texttt{TBabs} \citep{wilms2000absorption}. We fix the column density $N_{\rm H}$ of \texttt{TBabs} to $2.6\times10^{20}{\rm cm}^{-2}$, which is slightly larger than the $N_{\rm H}$ value used in previous studies ($\sim2.3\times10^{20}~{\rm cm}^{-2}$; e.g., \citealt{lin2017large,webbe2023variability}) derived from the density of the atomic hydrogen from 21~cm survey data \citep{kalberla2005leiden}. The new $N_{\rm H}=2.6\times10^{20}~{\rm cm}^{-2}$ is derived by mapping Galactic absorption using the X--ray afterglows of $\gamma$--ray bursts \citep{willingale2013calibration}, taking into account hydrogen in both atomic and molecular form\footnote{In \texttt{TBabs}, the fraction of molecular hydrogen present as H$_2$ is assumed to be 20\% \citep{wilms2000absorption}, similar to the measured fraction along the line--of--sight towards J1231 (18.5\%--20.3\%; \citealt{willingale2013calibration}).}. We present the tables of fit parameter constraints in Appendix.

\section{Results}

\subsection{Low black hole mass and high spin from slim disc modelling}
\label{sc:modelling}

\begin{figure*}
   \centering
   \includegraphics[width=0.45\textwidth]{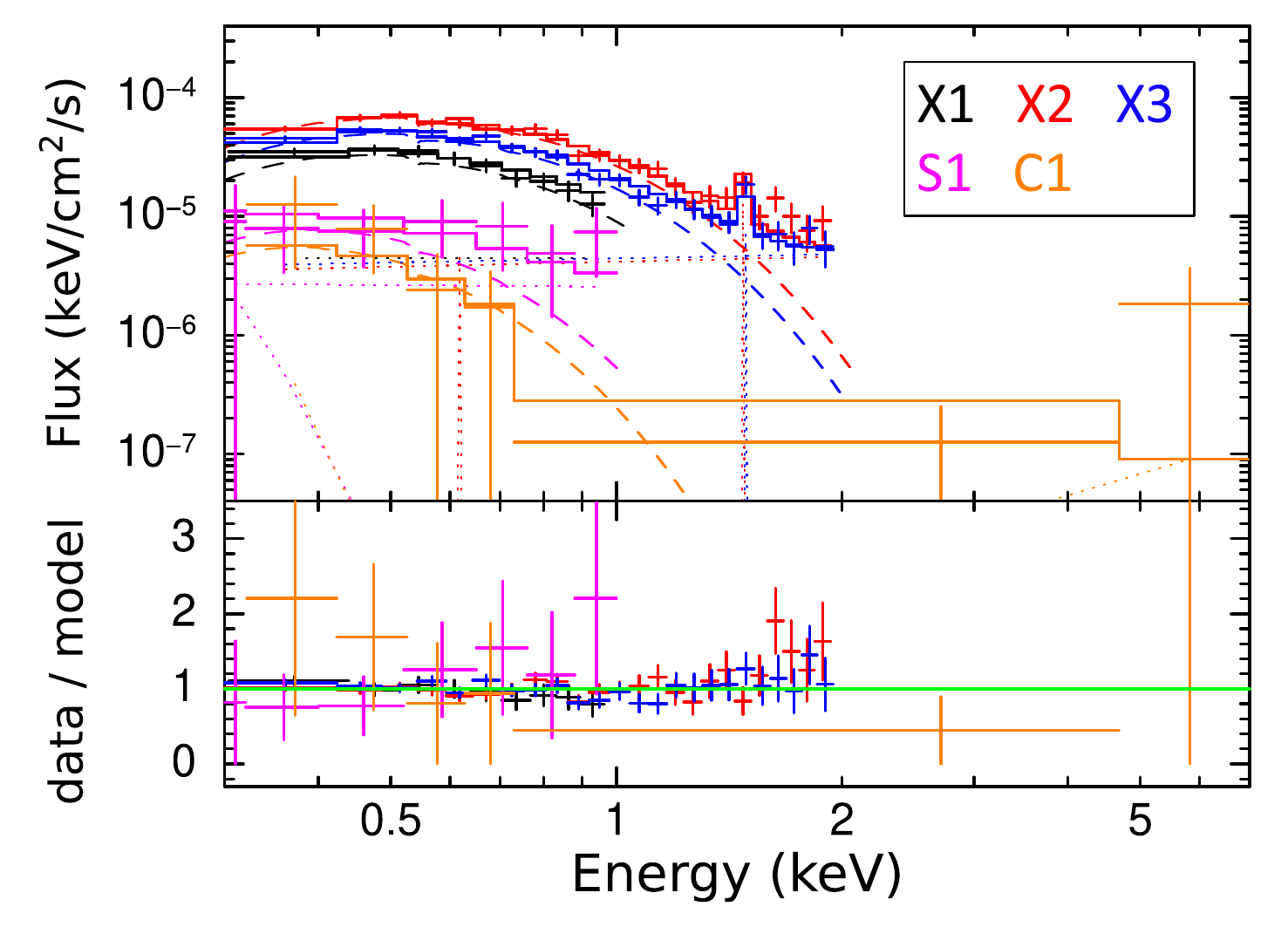}\hfill
   \includegraphics[width=0.45\textwidth]{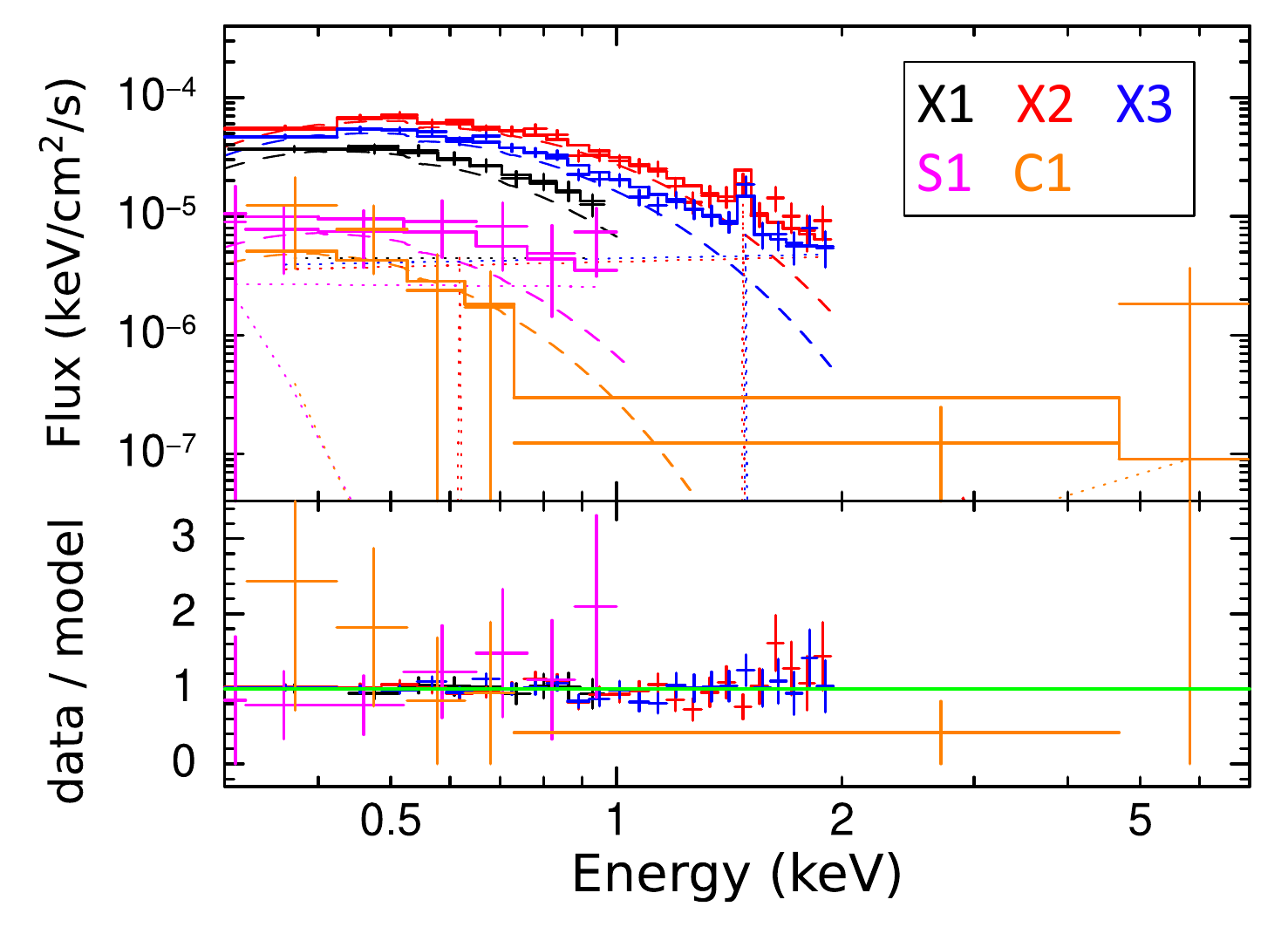}
   \caption{\textit{Left}: a joint fit of the J1231 data at five epochs (colour--coded) with a fit function \texttt{"constant*TBabs*slimdz"}. The EPIC/MOS data are not included for plotting purpose only. The solid, dashed, and dotted lines represent the total model, the slim disc, and the background components, respectively. The bottom panel shows the ratio between the observed number of counts in each spectral bin (data) and the best--fit predicted number of counts in each spectral bin (model; solid lines in the top panel). Here the spectral evolution is well explained by a slim disc with only its $\dot m$ varying between epochs (Table~\ref{tb:jointfit}). \textit{Right}: Same as the left plot, but here the spectral evolution between the three \xmm{} epochs is explained by a varying $\theta$ while the $\dot m$ remains a constant between those epochs.}
   \label{fig:jointfit}
\end{figure*}

\begin{figure}
   \centering
   \includegraphics[width=0.8\linewidth]{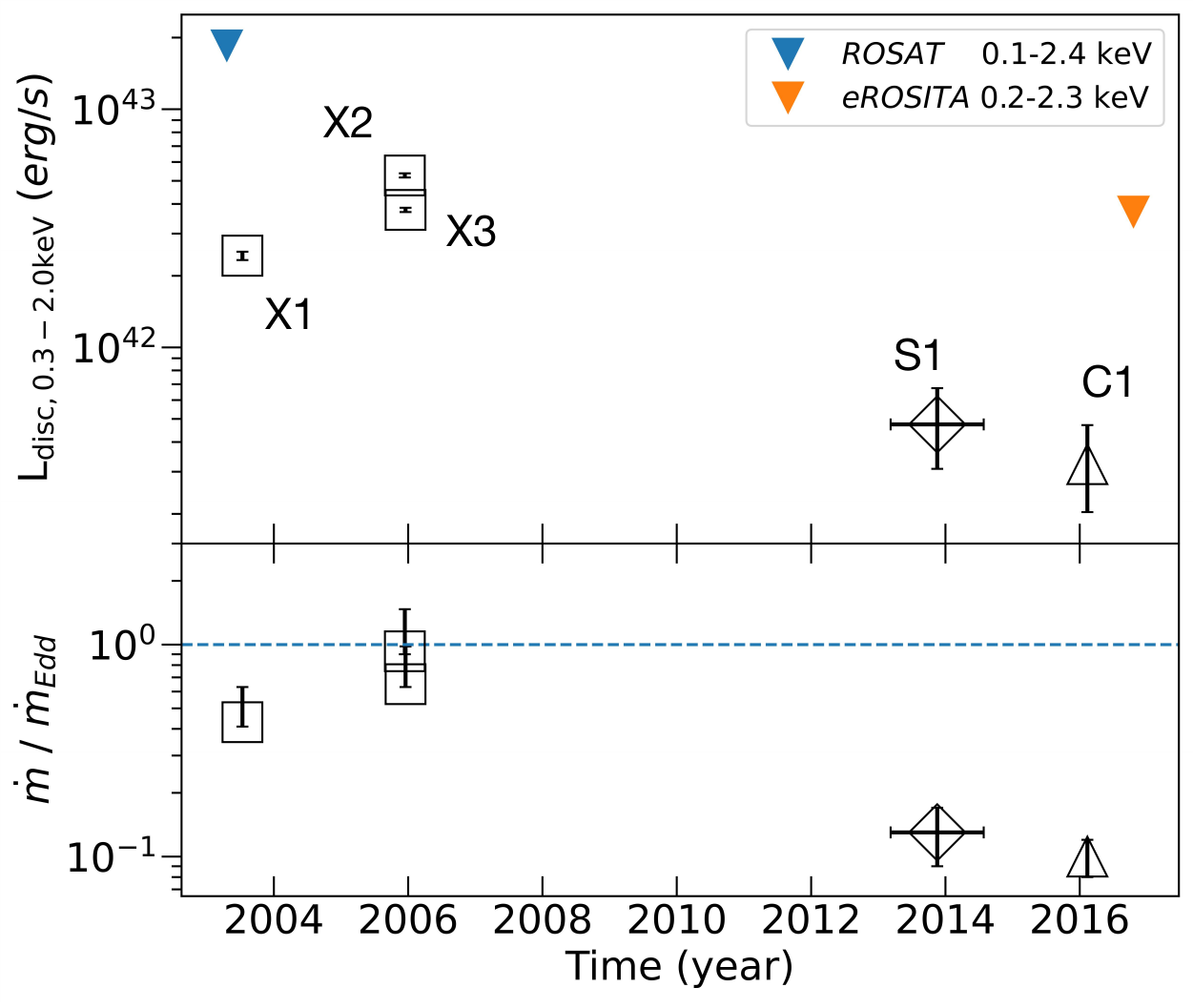}
   \caption{\textit{Top panel}: We plot the disc luminosity in the 0.3--2.0~keV band versus time as derived from our spectral analysis considering no disc inclination changes between all data epochs (Table~\ref{tb:jointfit}). The square, diamond, and triangle symbols represent the \xmm{}, \swf{}, and \chan{} observations, respectively. We also indicate the upper limits from \textit{ROSAT} and \textit{eROSITA} surveys (time shifted for plotting purpose; \citealt{voges1999rosat,tubin2024erosita}). \textit{Bottom panel}: the fitted disc mass accretion rate $\dot m$ associated with each epoch, normalized by the Eddington-limited accretion rate (computed assuming a BH mass of $5.7\times10^4$~$M_{\odot}$).  We indicate trans-Eddington accretion with a dashed line.}
   \label{fig:lum}
\end{figure}

\begin{figure*}
   \centering
   \includegraphics[width=0.38\textwidth]{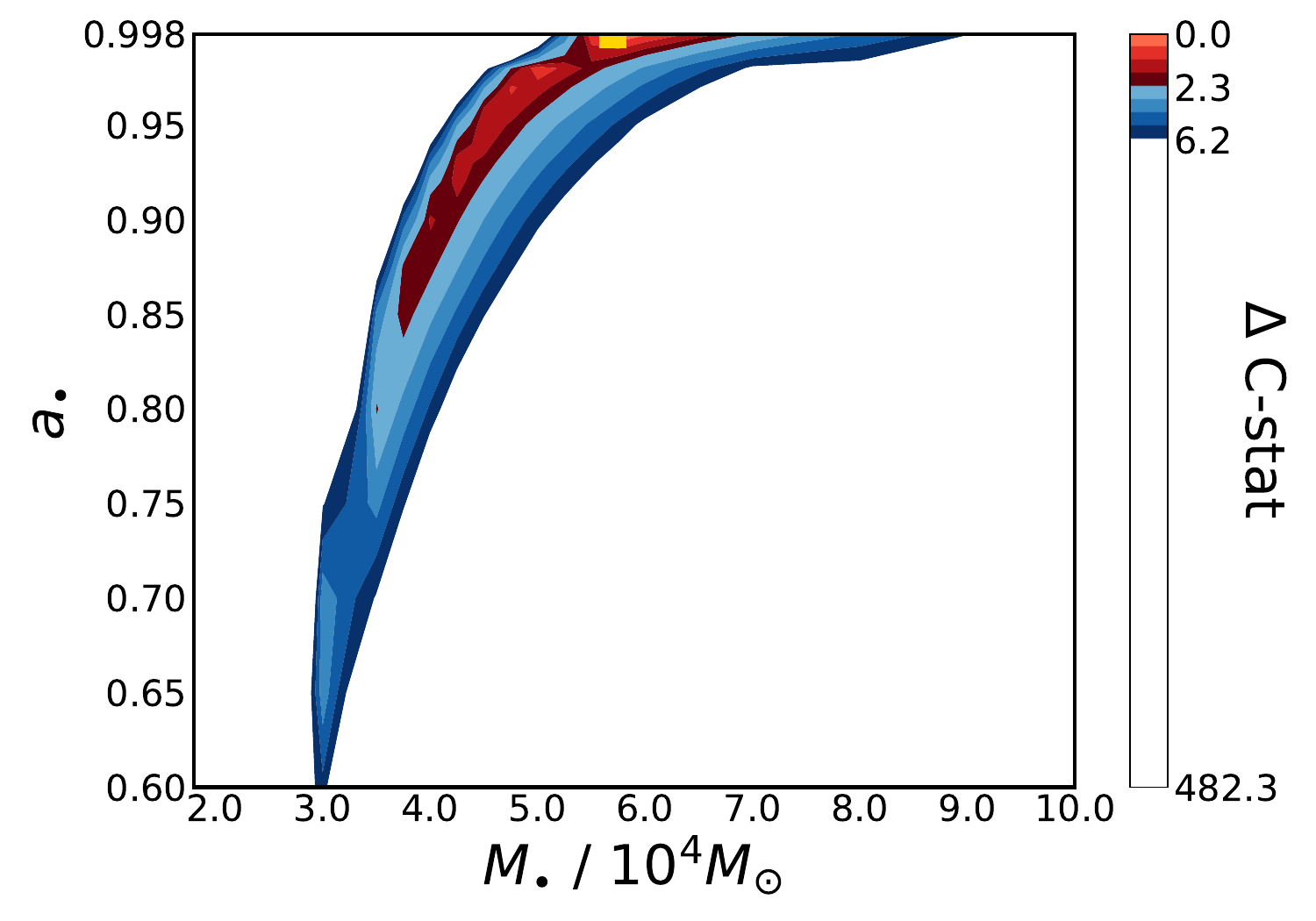}
   \includegraphics[width=0.38\textwidth]{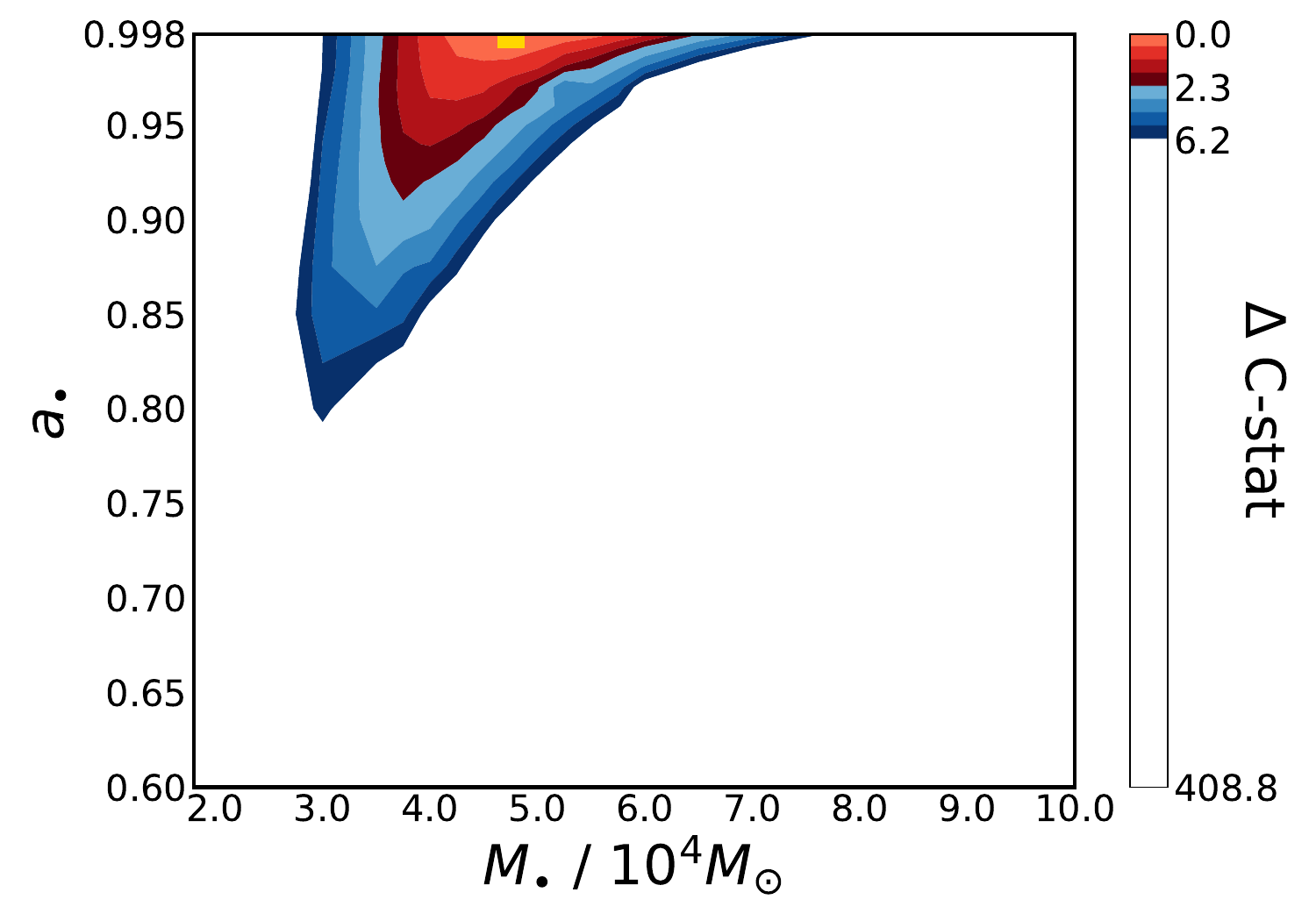}
   \caption{Constraints on $M_{\bullet}$ and $a_{\bullet}$ from the slim disc model--fit to the spectra obtained at all observing epochs. We calculate the $\Delta$C-stat across the $\{M_{\bullet}$, $a_{\bullet}\}$ plane, with respect to the best--fit value (yellow marker) from Table~\ref{tb:jointfit} (\textit{left}) and from Table~\ref{tb:jointfit-theta} (\textit{right}). Areas within 1$\sigma$ and 2$\sigma$ confidence limits for two--parameter error estimates are filled by red and blue colours, respectively. Considering both scenarios, at 2$\sigma$ for the two-parameter fits, $M_{\bullet}$ is constrained to be ($6\pm3)\times10^{4}$~M$_\odot$. The lower limit to the BH spin is constrained to be $>0.6$ at the 2$\sigma$ 97.8\% single--sided confidence level.}
   \label{fig:contour}
\end{figure*}

\begin{figure}
   \centering
   \includegraphics[width=0.8\linewidth]{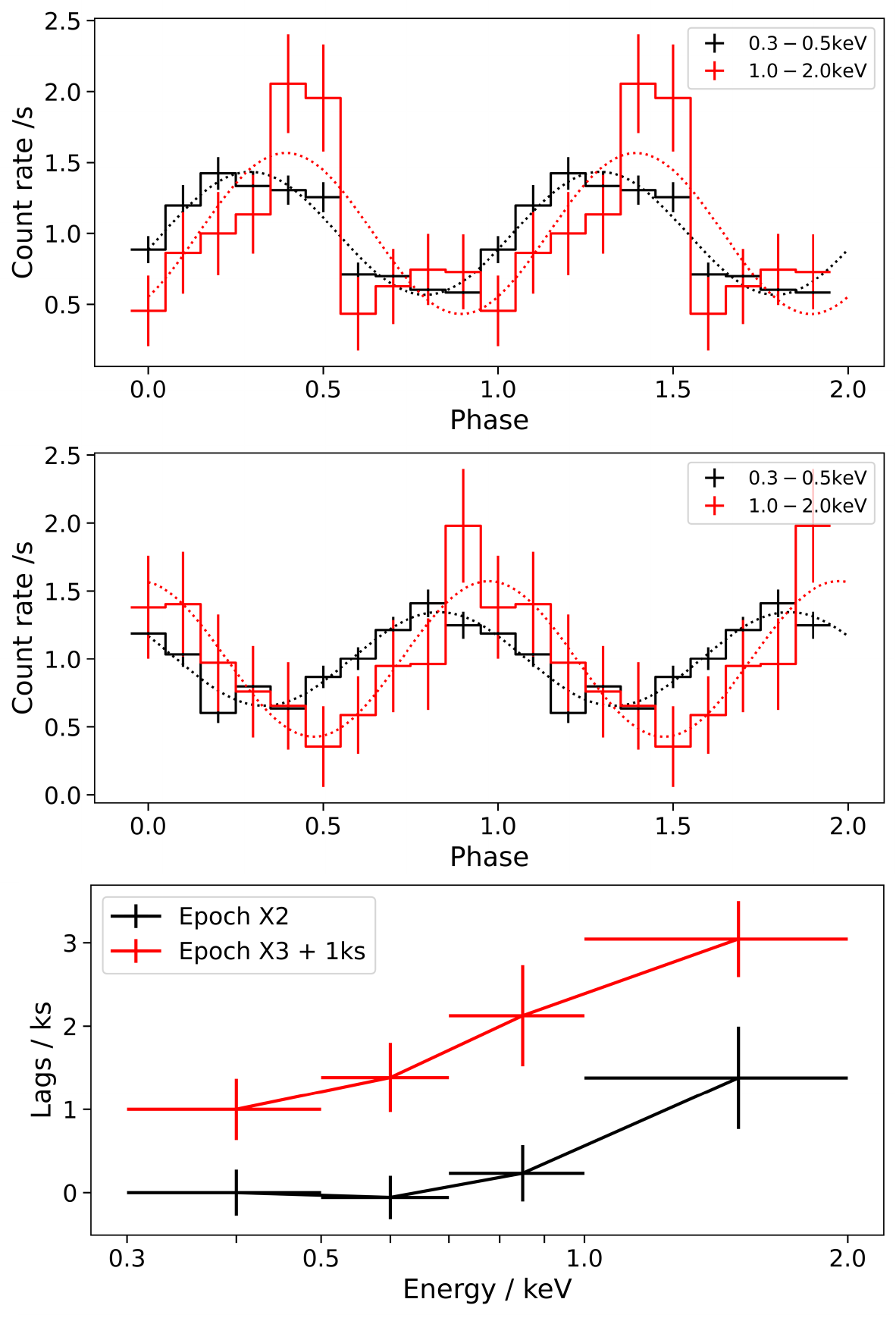}
   \caption{\textit{Top panel}: Phase--folded lightcurves of epoch X2. We extract the lightcurves at different energy bands: 0.3--0.5~keV, 0.5--0.7~keV, 0.7--1.0~keV, and 1.0--2.0~keV. For clarity, we only plot the 0.3--0.5~keV and 1.0--2.0~keV lightcurves here. Dotted lines are the sinusoidal function fit best to the data ($\chi^2/$degree--of--freedom$ <2$ for data at each energy band). \textit{Middle panel}: Phase--folded lightcurves of epoch X3. \textit{Bottom panel}: Time lags of the lightcurves in different energy bands with respect to the reference band of 0.3--0.5~keV. A positive lag means that the band of our interest lags the reference band in X--ray signals. In the plot, the points of X3 are shifted by +1~ks in time difference for clarity. The J1231 data shows that the hard band (1.0--2.0~keV) lags the soft band (0.3--0.5~keV) in time ($1.4\pm0.9$~ks at X2, and $2.0\pm0.8$~ks at X3). J1231 is thus unlike QPEs, which typically show a "hard--rise--soft--decay" mode, such that hard bands {\it lead} soft bands through each eruption.}
   \label{fig:foldlc}
\end{figure}

\begin{figure*}
   \centering
   \includegraphics[width=0.7\linewidth]{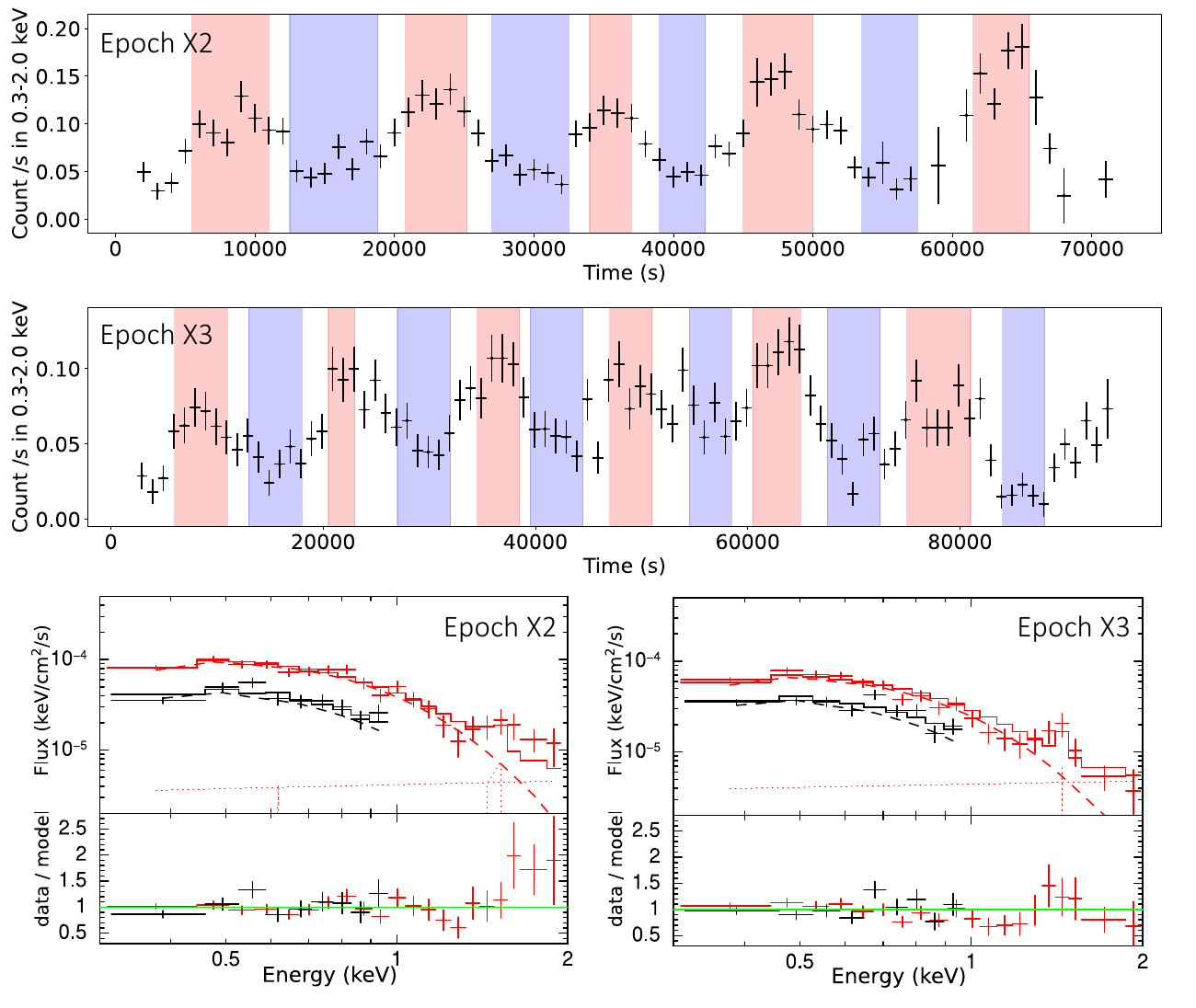}
   \caption{\textit{Top two panels}: The 0.3--2.0~keV light--curves in the X2 (\textit{top first}) and X3 (\textit{top second}) epochs, re--binned to 1000~s per time bin. We mark the time intervals selected for generating the spectra by red (peaks; X2--p and X3--p) and blue (valleys; X2--v and X3--v) colour. We manually choose the time intervals to select peaks and valleys. For each set of data we average the spectra. The un--selected data in white are not considered. The exact time intervals are listed in the Appendix. \textit{Bottom panels}: For each epoch, the black and red data are from the valley and the peak EPIC/pn, respectively. The EPIC/MOS data are not included for plotting purpose only. The solid, dashed, and dotted lines represent the total model, the slim disc, and the background components, respectively. Spectra at valleys have identical backgrounds as those at peaks. Therefore, the valley backgrounds are not shown. We freeze the best-fit slim disc model parameters (Table~\ref{tb:jointfit}) allowing only the $\dot m$ to vary to fit the peak and valley spectra. This approach yields good fits to the peak and valley spectra, see Table~\ref{tb:mdot} for the parameter constraints. Some fit residuals remain $\sim2$~keV where the background is dominant.}
   \label{fig:lcs}
\end{figure*}

We use the slim disc model \texttt{slimdz} \citep{wen2020continuum,wen2022library} to fit the accretion disc spectra of J1231, as its X--ray luminosity is close to Eddington when the disc advection should not be ignored. The slim disc
model considers the stationary, relativistic ``slim disc'' accretion disc solutions \citep{abramowicz1988slim} and ray--traces the disc photons self--consistently to the observer's frame. The model implements the astrophysical spin limit of a Kerr BH $-0.998<a_{\bullet}<0.998$ \citep[][]{thorne1974disk}. In the \texttt{slimdz} model, the disc radiative efficiency, $\eta$, is fixed to 0.1 only to determine the unit of the mass accretion rate: $\dot m_{\rm Edd}=1.37\times10^{21}$~kg~s$^{-1}(0.1/\eta)(M_{\bullet}/10^{6}M_{\odot})$. The actual disk radiative efficiency can vary between epochs (as expected in the slim--disc scenario when the $\dot m$ changes; e.g., \citealt{abramowicz2013foundations}), and can be determined from the physical value of $\dot m$ after constraining the mass $M_{\bullet}$ and spin $a_{\bullet}$. Further details of \texttt{slimdz}, including the assumption of a fixed viscosity $\alpha=0.1$, are presented in \citealt{wen2022library}. Comparisons of the \texttt{slimdz} model with other thin disc models (e.g., \texttt{optxagnf}; \citealt{done2012intrinsic}) have been made \citep[e.g.,][]{cao2023rapidly}.

We note that \texttt{slimdz} was originally designed for TDE sources in that it assumes an outer disc radius $R_{\rm out}\leq600$~$R_{\rm g}$ \citep{wen2022library}. However, even if J1231 is a variable AGN and not a TDE, the disc region $\geq600$~$R_g$ of a much larger ($>10^{3}$~$R_g$) AGN slim disc would contribute little to the X--ray spectrum ($\lesssim1\%$; \citealt{wen2021mass}) at $M_{\bullet}\sim10^{3}$--$10^{6}$~$M_{\odot}$. Therefore, the spectral fits with \texttt{slimdz} are physically self--consistent irrespective of whether or not J1231 is a TDE. 

We fit all the epochs from X1 to C1 together using a fit function \texttt{"constant*TBabs*slimdz"}. As we find no evidence for significant intrinsic absorption from our phenomenological fits (summarised in Appendix), we do not consider the intrinsic absorption in all of our analysis. First, we assume the black hole mass $M_{\bullet}$, spin $a_{\bullet}$, and inclination $\theta$ do not change over the $\approx$13~yr period during which the data was taken. For this joint fit (we simultaneously fit spectra from all epochs), we leave those parameters free to vary while forcing them to be constant between epochs. Due to the low number of photons detected at S1 and C1, we cannot constrain the re--normalisation factor between the EPIC/pn and the \swf{}/XRT ($C_{\rm XRT}$), and the factor between the EPIC/pn and the \chan{}/ACIS ($C_{\rm ACIS}$), when $\dot m$ in the \texttt{slimdz} model is left as a to--be--fit parameter at S1 and C1. Therefore, we leave the $\dot m$ as a free parameter in the fit, while we fix both $C_{\rm XRT}$ and $C_{\rm ACIS}$ to unity during the joint--fit. Through this, we effectively assume the different instruments are perfectly cross--calibrated (while in reality the estimated uncertainties in $C_{\rm XRT}$ is $\lesssim0.5$\%, and in $C_{\rm ACIS}$ is $\lesssim10$\%; e.g., \citealt{plucinsky2017snr}).

We present the best--joint--fit results in Table~\ref{tb:jointfit} and Fig.~\ref{fig:jointfit}. The long--term X--ray evolution of J1231, i.e., the decade--long decay, can be explained as due to variations in the mass accretion rate $\dot m$ through the disc. Specifically, the $\dot m$ increases from $\sim0.4~\dot m_{\rm Edd}$ at X1, to $\sim\dot m_{\rm Edd}$ at X2 and X3, before it decreases to $\sim0.1~\dot m_{\rm Edd}$ at later epochs. Fig.~\ref{fig:lum} summarises the long--term evolution of the disc 0.3--2.0~keV luminosity, as well as the $\dot m$, based on our analysis results. Fig.~\ref{fig:contour} shows the $\Delta$C-stat contours in $\{M_{\bullet}$, $a_{\bullet}\}$ space, where the BH mass is constrained to be $(6\pm3)\times10^4~M_{\odot}$ and the spin to be $>0.6$ at 2$\sigma$ (97.8\% single--sided for the spin value) confidence. Note that the errors are for two--parameter error estimates counting for dependencies between the constraints of $M_{\bullet}$ and $a_{\bullet}$, rather than the single--parameter error estimates in Table~\ref{tb:jointfit}.

We also test if the luminosity increase from X1 to X2 can be explained by the effect of a varying disc inclination $\theta$, instead of the $\dot m$ variation. In this scenario, a TDE disc is mis--aligned to the BH equatorial plane initially. It is possible to take $\sim10^3$ days before the disc is eventually aligned to the BH equatorial plane \citep[e.g.,][]{franchini2016lense}. Therefore, in this test fit we assume that the $\theta$ at X1/X2/X3 epochs are free to vary from one epoch to another. At S1 and C1, $\gtrsim10$~yr after the first detection, we assume the disc alignment is terminated and the $\theta$ stays the same between these two epochs, while this late--time $\theta$ is a to--be--fit parameter in the fit--function as well, since the $\theta$ of the aligned disc can be different from that of a disc during the alignment process at earlier epochs.

We find that the spectral evolution from X1 to X3 can be well--explained by a disc with a fixed $\dot m$ and a varying $\theta$ (Table~\ref{tb:jointfit-theta}). Letting $\dot m$ free to vary between X1/X2/X3 will not improve the fit significantly (from C-stat/d.o.f.$=$119.0/123 to 118.0/121). In this joint--fit, although the marginal confidence ranges of $\theta$ at X1/X2/X3 overlap with each other, we confirm that for any given set of \{$\dot m$, $M_{\bullet}$, $a_{\bullet}$\}, from X1 to X3 the $\theta$ are different so to account for the spectral differences between them. In Fig.~\ref{fig:contour}, we also present the $\Delta$C-stat contours in $\{M_{\bullet}$, $a_{\bullet}\}$ space derived from this joint--fit, to compare it with the results from the previous joint--fit with a varying $\dot m$ between X1/X2/X3. Here the BH mass is constrained to be $(5\pm2)\times10^4~M_{\odot}$ and the spin to be $>0.8$ at 2$\sigma$ (97.8\% single--sided for the spin value) confidence.

From the fit statistics (C-stat), it seems that this scenario of an early--time $\theta$ variation describe the data better than the previous scenario of no $\theta$ changes between all data epochs (C-stat/d.o.f.$=119.0/123$ compared to 136.9/124). However, based on the polynomial expressions given by \citet{kaastra2017use}, we calculate the expected value of C-stat, $C_e\approx135$, with an RMS of $\approx16$. We also check the spectral fits in the data--model ratio plot (Fig.~\ref{fig:jointfit}) and see no significant differences between the two fits. No signs of over--fitting are found. Therefore, in this paper we consider both scenarios good descriptions of the data and our main results. We should note that, although the data is sufficiently explained by either $\dot m$ or $\theta$ variation between epochs, the scenario that both parameters are varying over time is not excluded.

The spectra at S1 and C1 are likely aligned--disc spectra regardless of various scenarios explaining the source behaviour at earlier epochs (X1 to X3). By fitting only these two spectra, however, we test and find that they alone do not distinguish between different scenarios due to low photon counts. In such a way, we also check that our choices of $C_{\rm XRT}=1$ and $C_{\rm ACIS}=1$ during the joint--fits (Tabel~\ref{tb:jointfit} or Tabel~\ref{tb:jointfit-theta}) do not result in BH mass and spin values that are unable to explain the spectra at S1 or C1.

\subsection{An unusual short-term variability pattern for a QPE}

It has been proposed that J1231's $\sim3.8$~hr variability could be due to a QPE. A known QPE variability pattern is a "hard--rise--soft--decay" mode, which manifests itself as the hard X--ray flux peaks before the soft flux does through each eruption \citep[e.g.,][]{miniutti2019nine,arcodia2024more,giustini2024fragments}. To check if J1231 behaves similarly to QPE sources with such spectral evolution, we examine the lightcurves at four energy bands: 0.3--0.5~keV, 0.5--0.7~keV, 0.7--1.0~keV, and 1.0--2.0~keV. We find no clear evidence for the peaks of hard bands (0.7--1.0~keV or 1.0--2.0~keV) leading those of the soft band (0.3--0.5~keV) in time. Fig.~\ref{fig:foldlc} shows the phase--folded EPIC/pn lightcurves at epoch X2 and X3. The periods for X2 and X3 are taken to be 13.52~ks and 14.35~ks, respectively \citep{webbe2023variability}. When phase-folding the lightcurves of epoch X2 and X3, we set the start time (T=0) to be at 2.50856724E+08~s and at 2.51185225E+08~s, respectively. Here the start time is given in offsets in seconds from the \xmm{} Mission Reference Time (MRT, 1997-12-31T23:58:56.816 UTC).

To quantify the time lags, we fit the phase--folded lightcurves with a sinusoidal fit function $CR_{i}=A_i{\rm sin}[2\pi(x-l_i)]+CR_{0,i}$, where $CR_{i}$ is the count rate at energy band $i$, $A_i$ is the amplitude, $x$ is the phase value from the x--axis, $l_i$ is the phase shift, and $CR_{0,i}$ is the count rate at phase $x=l_i$. The phase lag $\Delta$$l$ is defined as the change in the $l$ value between the energy band $i$ of our interest and the reference band. We then infer the time lags between peaks at different energy bands by multiplying the phase lags and the periods. In this way, a positive time lag means that the band of our interest lags the reference band in X--ray signals. The time lag--energy plot (Fig.~\ref{fig:foldlc}; reference band 0.3--0.5~keV) shows that, during the short--term variability in both epoch X2 and X3, the 1.0--2.0~keV band lags the 0.3--0.5~keV band in time, by $1.4\pm0.9$~ks at X2, and by $2.0\pm0.8$~ks at X3. This lag result is contrary to a "hard--rise--soft--decay" eruption (that the hard bands should lead the soft bands), suggesting that J1231's X-ray spectral variability is unlike that of typical QPEs during flares.

\subsection{Short--term variability from a changing disc mass accretion rate or a changing disc inclination?}
\label{sc:mdot}

We also fit the slim disc model to the time--resolved spectra within X2 and within X3, to test if the quasi--periodic variability at X2 or X3 introduces changes in the spectral shape within a single observation. Based on the \xmm{}/EPIC-pn lightcurves, we select spectra according to the 0.3--2.0~keV count rate during local lightcurve maxima and minima, and then average those spectra to create a peak and a valley composite, respectively. Due to the lightcurve evolution between cycles, it is not possible to have strict count rate criteria for a peak and a valley across different cycles. Also, as the variability is not strictly periodic, we cannot  separate the peaks and valleys based on the phase change of a fixed period to create the phase--resolved spectra. Therefore, we manually choose time intervals for peaks and valleys, and we discard the data in between. Next, we employ the SAS command {\sc gtibuild} to combine the data of the selected time intervals and create the average peak and valley spectra from both pn and MOS for both epoch X2 and X3. We refer to the peak spectra as X2--p and X3--p, and to the valley spectra as X2--v and X3--v. Due to the decrease of source count rate below that of the background for energy bins $>1$~keV, we only consider the energy range 0.3--1.0~keV for the X2--v and X3--v spectra in the analysis. Fig.~\ref{fig:lcs} shows the time intervals selected during X2 and X3, which we also list in the Appendix. 

At each epoch, we fit a composite spectrum made from averaging the spectra at the peaks of the short-term variability and a composite spectrum from averaging the valleys. Our multi--epoch disc modelling in Section~\ref{sc:modelling} reveals that at X2 \& X3, the ``average'' values of \{$\dot m$, $\theta$\} can either be both high ($\sim10\dot m_{\rm Edd}$, $\sim70^{\circ}$), or both low ($\lesssim\dot m_{\rm Edd}$, $<30^{\circ}$). Here, for each scenario, we test if the spectral difference between the peaks and valleys can be explained as due to a single changing parameter associated with the disc ($\dot m$ or $\theta$).

For the scenario of low average $\theta$ at X2 \& X3, we fix the values of $M_{\bullet}$ and $a_{\bullet}$ to the best--fit values from Table~\ref{tb:jointfit}. Within both X2 and X3, the spectral difference between the peaks and valleys can be explained as due to a changing mass accretion rate (Fig.~\ref{fig:lcs} and Table~\ref{tb:mdot}), while $\theta$ is fixed to $10^{\circ}$. Here $C_e$ is $\approx$68 with an RMS of $\approx$12 for both epochs. Meanwhile, the model is rejected by the data at 99\% confidence level (C-stat$>C_e+2.33\times$RMS) if we force the $\dot m$ to be the same ($\dot m=0.93$ for X2 and $\dot m=0.65$ for X3) and letting $\theta$ change between peaks and valleys. Therefore, if the system has a low average $\theta$ at X2 \& X3, the data is consistent with a varying $\dot m$ explaining the short--term spectral variability at X2 \& X3. This $\dot m$ conclusion holds qualitatively when we assume the case of the smallest spin value based on results in Fig.~\ref{fig:contour} ($M_{\bullet}=3\times10^4$~$M_{\odot}$ and $a_{\bullet}=0.6$; see Table~\ref{tb:smallspin}).

For the scenario of high average $\theta$, similar to the low--$\theta$ case above, we fix the values of $M_{\bullet}$ and $a_{\bullet}$ to the best--fit values from Table~\ref{tb:jointfit-theta} and test if the spectral difference between peaks and valleys can be explained by a single $\dot m$ or $\theta$ parameter. The model is rejected by the data at one or both epochs at 99\% and 95\% confidence level (C-stat$>C_e+1.645\times$RMS), for the varying--$\dot m$ case and the varying--$\theta$ case, respectively. We note that a good fit with large uncertainties can be achieved by allowing both $\dot m$ and $\theta$ to vary between peaks and valleys (and Table~\ref{tb:highincl}).

\section{Discussion}
\label{discuss}

\subsection{A rapidly--spinning IMBH}

Despite different scenarios explaining the spectral evolution, we constrain the BH mass of J1231 to be $(6\pm3)\times10^4$~$M_{\odot}$ (Fig.~\ref{fig:contour}), similar to the value estimated from extrapolating the scaling relation for BH mass and optically-derived host galaxy velocity dispersion: $\sim10^5$~$M_{\odot}$ \citep{ho2012low}. Depending on different assumptions of $a_{\bullet}$ and the corona presence, studies using the AGN--like, thin--disc model have found the $M_{\bullet}$ constraints ranging from $\sim4\times10^{4}$ to $\sim2\times10^{6}$~$M_{\odot}$ \citep{lin20133,lin2017large}. Our study more strongly suggests that the central engine of J1231 is an IMBH, while counting the significant disc advection in Eddington and super--Eddington discs (especially for the super--Eddington scenario with a varying $\theta$; Table~\ref{tb:jointfit-theta}).

Furthermore, our disc modelling indicates the BH has a high spin: $a_{\bullet}>0.6$. A highly--spinning IMBH of $\lesssim10^5$~$M_{\odot}$ could be formed via direct collapse of a gas cloud in the early Universe \citep[e.g.,][]{loeb1994collapse,bromm2003formation}. Simulations have shown that $a_{\bullet}>0.9$ can be produced if the collapsing cloud goes through a supermassive stellar (SMS) phase before collapsing into a BH \citep[e.g.,][]{reisswig2013formation,inayoshi2014formation}. Alternatively, if the BH was born at a much lower mass ($\lesssim10^3$~$M_{\odot}$; e.g., via stellar remnants or gravitational runaway stellar collisions \citealt{madau2001massive,zwart2002runaway,devecchi2009formation,greif2011simulations}), then it must have gained its last $e$-fold in mass through subsequent accretion episodes. In such cases,  reaching a high spin while avoiding being spun--down due to multiple accretion episodes \citep[e.g.,][]{king2008evolution,metzger2016wind} requires that the seed BH grew to its current mass in one or more accretion episodes where the angular momentum vector of the accreted material was aligned with that of the BH spin. 

\subsection{Long--term X--ray evolution explained by disc changes}

Fig~\ref{fig:lum} shows the long--term evolution of the J1231 disc luminosity in the 0.3--10~keV band, as well as the $\dot m$ evolution as derived from our analysis of the X-ray spectra based on the scenario of no $\theta$ changes between all data epochs. In this scenario, we find that the $\dot m$ of the disc increases to $\approx\dot m_{\rm Edd}$ at X2 and X3 compared to $\approx0.4\dot m_{\rm Edd}$ at X1 two years earlier, before it drops to $\sim0.1$~$\dot m_{\rm Edd}$ at S1 and C1 several years later (Table~\ref{tb:jointfit}). Assuming a linear interpolation between epochs in Fig~\ref{fig:lum}, we roughly estimate the total mass accreted between 2004 and 2016 to be $\sim0.01$~$M_{\odot}$. The small amount of total accreted mass suggests that, if J1231 is a bona--fide TDE \citep[e.g.,][]{rees1988tidal,strubbe2009optical,metzger2016bright}, it might be either a weak partial disruption that stripped off very little mass, or a full disruption of a subsolar object like a brown dwarf or large gas giant.

Regardless of whether J1231 is a TDE or solely due to an active nucleus, in this scenario (Table~\ref{tb:jointfit}), $\dot m$ at all epochs lies in the range $\dot m_{\rm Edd}\gtrsim\dot m\gtrsim0.1\dot m_{\rm Edd}$ (Table~\ref{tb:jointfit}). Classical disc theories predict that a disc radiation--pressure instability occurs in this $\dot m$ range (e.g., \citealt{lightman1974black,shakura1976theory,piran1978role}; see \citealt{czerny2019slim} for a review), preventing a steady $\dot m$ in this instability range and forcing the disc to go through the so--called ``limit cycle'' \citep[e.g.,][]{lasota11991variability,szuszkiewicz1998limit,xue2011studies}. In such case the $\dot m$ should avoid values in the instability range over a timescale larger than the thermal timescale at the outer edge of disc instability zone ($\sim$days in the J1231 case).

However, observational evidence from the XRB population indicate that a steady disc with $\dot m$ in the $\dot m_{\rm Edd}\gtrsim\dot m\gtrsim0.1\dot m_{\rm Edd}$ instability range, remains likely. Except for two XRBs (GRS~1915+105 and IGR~J17091-3624; e.g., \citealt{belloni1997unified,janiuk2000radiation,janiuk2015interplay,altamirano2011faint}), most XRB discs in the range $\dot m_{\rm Edd}\gtrsim\dot m\gtrsim0.1\dot m_{\rm Edd}$ do not have signs of radiation--pressure instability \citep[e.g.,][]{gierlinski2004black,czerny2019slim}. It is purposed that other factors such as magnetic fields may stabilise the accretion disc to prevent a limit cycle from occurring \citep[e.g.,][]{janiuk2011different,kaur2023magnetically}. Strong outflows when the disc accretion is at high--Eddington/super--Eddington levels \citep[e.g.,][]{middleton2013broad,pinto2016resolved,pinto2021xmm,kara2018ultrafast,pasham2024case} will also help to stabilise the disc. These stabilising mechanisms could be important in J1231, as we find that at all data epochs and for constant $\theta$ across all epochs, the source spectrum is always consistent with a steady disc model of $\dot m$ in the range $\dot m_{\rm Edd}\gtrsim\dot m\gtrsim0.1\dot m_{\rm Edd}$. Meanwhile, as long as the disc instability is avoided, the slim disc model remains applicable to the S1 and C1 spectra at $\sim0.1\dot m_{\rm Edd}$, because the slim disc and thin disc share the same branch of disc solutions, and the additional terms in the slim disc model automatically become insignificant if $\dot m<<\dot m_{\rm Edd}$ \citep[e.g.,][]{abramowicz2013foundations,czerny2019slim}.

Several scenarios have been suggested by \citet{lin2017large} to explain the luminosity increase from X1 to X2, for instance, the X1 epoch catching the initial fast rise of the TDE disc, or a slow disc circularisation in the TDE (e.g., \citealt{guillochon2015dark,hayasaki2021origin}), or a prolonged disruption of an evolved star (\citealt{macleod2012tidal}; though this TDE subclass is typically of low likelihood, see e.g., \citealt{macleod2013spoon,kochanek2016tidal}). In our study we find that the source spectrum is always consistent with a disc spectrum starting from X1, results in line with the scenarios except a slow disc circulariation process. 

Could Lense--Thirring precession, arising from a misalignment between the BH's equatorial spin plane and the disc plane shortly after the TDE \citep[e.g.,][]{stone2012observing}, be responsible for J1231's long--term variability? Observational evidence supports such a scenario in two TDEs (ASASSN-14li; \citealt{pasham2019loud}; AT2020ocn; \citealt{pasham2024lense,cao2024tidal}). In the case of J1231, if a solar--type star is disrupted, the predicted precession period for a $<10^5$~$M_{\odot}$ BH is $\gtrsim1$~day, irrespective of the value of the BH spin \citep[e.g.,][]{franchini2016lense,teboul2023unified} (though we note that the precession period scales $\propto$~$R_{\rm out}^{3}$ at first order; e.g., \citealt{pasham2024lense}). We test and find that the luminosity changes from X1 to X3 can be well explained by a mis--aligned disc going through the alignment process while the $\dot m$ stays at super--Eddington (Tabel~\ref{tb:jointfit-theta}). In this scenario, the $\theta$ variation during the disc alignment leads to the luminosity variation, which is sparsely sampled by three early \xmm{} observations. It is also possible that, the QPOs are not observed at X1 but only at X2 and X3, because they are obscured by the edge of the slim disc, due to the $\theta$ becoming larger at X1 than at X2 or X3.

Meanwhile, \citet{wen2020continuum} find a TDE observed nearly edge--on would also result in a slow luminosity rise after the disruption, due to the process they called disc slimming. Our results suggest either the disc is viewed face--on (Table~\ref{tb:jointfit}), or it is the varying $\theta$ that causes the luminosity change (Table~\ref{tb:jointfit-theta}), disfavouring the disc slimming scenario in J1231. A delayed X--ray luminosity increase with respect to the initial disruption can also be explained by a partial TDE where the star is not fully disrupted during its first passage through the pericenter, allowing subsequent disruptions and mass accretion \citep[e.g.,][]{wevers2023live,liu2023deciphering}. Alternatively, J1231 might not be a TDE, but a variable AGN. However, in this case J1231 would be atypical for AGNs to have pure thermal X--ray spectra, while only 1.5\% AGNs vary in X--rays by a factor of >10 \citep{lin2012classification,lin2017large}.

\subsection{QPOs at X2 \& X3 does not show the typical QPE mode}

Besides the long--term spectral evolution over a decade, J1231 shows a short--term X--ray variability at X2 and X3, on the timescale of $\sim3.8$~hr ($\approx$0.07~mHz). \citet{lin20133} first reported the variability, and they proposed it to be analogous to the low--frequency QPOs (LFQPO) detected in X--ray binaries (XRBs) which host a stellar--mass BH. Assuming a linear anti-correlation of the LFQPO frequency with BH mass, a 0.07~mHz QPO in an accretion disc around a $6\times10^4$~$M_{\odot}$ BH corresponds to $\sim0.1$~Hz for a 10~$M_{\odot}$ BH, reminiscent of the lower--frequency end of LFQPOs in XRBs (0.1--30~Hz; \citealt{belloni2002unified,remillard2006x}). Moreover, the disappearance of the QPO features in J1231 (when the X--ray luminosity decreases at later epochs) resembles the spectral state transition between the ultra-luminous state (ULS; also named as the steep powerlaw state) to the thermal state in XRBs \citep[e.g.,][]{remillard2006x,li2014peak}. We note that factors other than the BH mass (e.g., disc accretion rate) might also impact the LFQPO frequency \citep[e.g.,][]{mchardy2006active,li2014peak,van2020centroid}.

The multi--epoch spectral analysis in Section~\ref{sc:modelling} reveals that at X2 \& X3, the ``average'' $\theta$ is either $<30^{\circ}$ or $\sim70^{\circ}$, depending on whether we assume a constant $\theta$ between all data epochs (Table~\ref{tb:jointfit} and Table~\ref{tb:jointfit-theta}). For each scenario, we test on the peak and the valley spectra to see if a single disc parameter ($\dot m$ or $\theta$) could explain the short--term variability. For a face--on disc ($\theta<30^{\circ}$), we find a difference in $\dot m$ explains the short--term spectral evolution, while the model varying only $\theta$ fails to explain the data; for a large $\theta\sim70^{\circ}$, however, the peak and the valley spectra can only be simultaneously explained if both $\theta$ and $\dot m$ are varying between the peaks and the valleys (see Section~\ref{sc:mdot} for details of the tests). We note that these results do not exclude the large--$\theta$ scenario (i.e., Table~\ref{tb:jointfit-theta}) as the explanation for J1231's long--term X--rays, as the short--term variability could possibly be caused by mechanisms beyond the description of a quasi--stable slim disc (e.g., a localised disc region causing the variability, as proposed to explain QPOs; \citealt{tagger1999accretion,chakrabarti2000correlation}).

It has also been proposed that J1231's short--term variability is a variant of the QPEs seen in some TDEs and AGN \citep[e.g.,][]{miniutti2019nine,giustini2020x,arcodia2021x,chakraborty2021possible,evans2023monthly,quintin2023tormund,webbe2023variability,nicholl2024quasi,arcodia2024more,guolo2024x}. If the quasi--periodic signal at X2 and X3 is indeed due to QPEs, our results provide evidence of an $\dot m$ variation as the driver of the QPE phenomenon. Alternatively, the accretion disc might not be circularised or thermalised during a QPE \citep{krolik2022quasiperiodic,king2023qpe}, so that the \texttt{slimdz} model is not applicable to X2 and X3 (though we find the source is consistent with a disc spectrum at the valleys and peaks). However, we like to stress that the J1231 spectral variation at X2 and X3 differs from that observed in QPE sources. Contrary to the "hard--rise--soft--decay" eruptions in typical QPE sources (e.g., GSN~069; \citealt{miniutti2019nine}), in the phase--folded lightcurves we find that the hard energy band (1.0--2.0~keV) lags the soft band (0.3--0.5~keV), resulting in a "soft--rise--hard--decay" mode (Fig.~\ref{fig:foldlc}). One possibility to explain the hard lag is that the $\dot m$ variation propagates from the outer disc, where most of the soft photons come from, to the innermost disc region, where most of the harder photons come from. This propagation takes time and thus leads to a delayed hard band variation. As the nature of QPEs is still under active debate, however, it remains unclear if J1231 is a variant of the QPE phenomenon or not. 

\section{Conclusions}

We present our spectral analysis of J1231's evolving X--ray data, which are taken over more than a decade, from 2003 to 2016. Using a slim disc model for the accretion disc around the black hole, we find that the decade--long spectral evolution of J1231 can be explained with a varying mass accretion rate. Alternatively, a varying disc inclination with a constant disc accretion rate can well explain the spectral evolution between the three \xmm{} epochs within the first three years of the source's detection. A mis--aligned disc initially formed after the disruption event could possibly lead to this inclination variation. The best slim disc fit simultaneously to all the spectra yields a BH mass of ($6\pm3)\times10^{4}$~M$_\odot$ at 2$\sigma$ confidence, making J1231 
one of only a handful of intermediate--mass black hole candidates in the range $10^{2}$---$10^{5}$~M$_\odot$. The black hole spin is $>0.6$ at the 2$\sigma$ 97.8\% single--sided confidence level, a rapid spin. The source spectra $\gtrsim$10 years after the first observation are consistent with the same BH mass and spin.

Previous studies have found a short--term QPO with a $\sim3.8$~hr period in the X-ray light curve of J1231 when the source luminosity peaks, i.e., during the second and third of the five epochs observed. We separate the J1231 lightcurves into four different energy bands during the two QPO epochs. We find that the 1.0--2.0~keV hard band lags the 0.3--0.5~keV soft band by $\sim$kiloseconds, resulting in a "soft--rise--hard--decay" variation mode. This mode is in contrast to the typical QPE mode of "hard--rise--soft--decay", suggesting that J1231 is an atypical QPE candidate.

Furthermore, for each of those two epochs, we produce a composite spectrum of the QPO peaks and one of the valleys. Our analysis of the peak and valley composite spectra suggests that the QPO behaviour might as well be driven by a varying disc accretion rate, though other possibilities are not excluded. Such an accretion rate variation could be caused by any one of the mechanisms proposed also for QPEs. The hard--to--soft lag is explained by the inward propagation of the accretion rate variation on the disc. 

\begin{acknowledgements}
We thank the referee for comments that helped to improve this manuscript. This work made use of data supplied by the UK Swift Science Data Centre at the University of Leicester. This work used the Dutch national e-infrastructure with the support of the SURF Cooperative using grant no.~EINF-6770. PGJ is supported by the European Union (ERC, StarStruck, 101095973). Views and opinions expressed are however those of the author(s) only and do not necessarily reflect those of the European Union or the European Research Council.
AIZ acknowledges support in part from grant NASA ADAP \#80NSSC21K0988.
\end{acknowledgements}

%%%%%%%%%%%%%%%%%%%%%%%%%%%%%%%%%%%%%%%%%%%%%%%%%%
\section*{Data Availability}

All the X--ray data in this paper are publicly available from the data archive of HEASARC (https://heasarc.gsfc.nasa.gov/). A reproduction package is available at DOI: 10.5281/zenodo.14422714.

%%%%%%%%%%%%%%%%%%%% REFERENCES %%%%%%%%%%%%%%%%%%

% The best way to enter references is to use BibTeX:
\bibliographystyle{aa} % style aa.bst
\bibliography{reference}  % if your bibtex file is called example.bib

% Alternatively you could enter them by hand, like this:
% This method is tedious and prone to error if you have lots of references
%\begin{thebibliography}{99}
%\bibitem[\protect\citeauthoryear{Author}{2012}]{Author2012}
%Author A.~N., 2013, Journal of Improbable Astronomy, 1, 1
%\bibitem[\protect\citeauthoryear{Others}{2013}]{Others2013}
%Others S., 2012, Journal of Interesting Stuff, 17, 198
%\end{thebibliography}

%%%%%%%%%%%%%%%%%%%%%%%%%%%%%%%%%%%%%%%%%%%%%%%%%%

%%%%%%%%%%%%%%%%% APPENDICES %%%%%%%%%%%%%%%%%%%%%
\newpage
\begin{appendix}

\section{X--ray data reduction}

\subsection{XMM-{\it Newton} observations}
\label{sc:xmmdata}
\xmm ~observed J1231 over one epoch in 2003 and two in 2005. To perform the \xmm{} data reduction and extract the scientific products, we use the HEASOFT (version 6.33.2) and SAS (version 21.0.0) software packages with the calibration files released on April 23, 2024 (CCF release: XMM-CCF-REL-411). The source is outside the field--of--view\footnote{http://www.cosmos.esa.int/web/xmm-newton/mos1-ccd6} in one of the Metal Oxide Semi-conductor (MOS) cameras, MOS1, at Epoch X2 and X3. Meanwhile, the signal--to--noise ratio in the Reflection Grating Spectrometer (RGS) detectors is too low to perform spectral analysis. For consistency, in this paper we use only data from the pn and the MOS2 cameras (both are European Photon Imaging Cameras; EPICs). Therefore, we refer to MOS2 as MOS hereafter. 

We use the SAS command {\sc epproc} and {\sc emproc} to process the Science 0 data from the pn and the MOS camera, respectively. We exclude the data from periods with an enhanced background count rate, applying the standard filtering criteria\footnote{https://www.cosmos.esa.int/web/xmm-newton/sas-thread-epic-filterbackground} to each camera. We require that the 10--12~keV detection rate of pattern 0 events is $<$0.4 counts~s$^{-1}$ for the pn camera, and the $>$10~keV detection rate of pattern 0 events is $<$0.35 counts~s$^{-1}$ for MOS. The first of the two X1 data segments (exposure $\lesssim$20~ks) is discarded due to the presence of strong background flares. We extract data of the source at RA=12h31m03.24s, Dec=+11\degr06\arcmin48.6\arcsec~using circular regions centred on the source of 30\arcsec{} and 45\arcsec{} radii, for the pn and MOS cameras, respectively. These regions are larger than the circular source region of 25\arcsec{} used in \citealt{lin2013classification}, and they encircle the $\gtrsim90\%$ energy fraction at 1.0~keV for an off--axis ($\sim$7\arcmin{}) point source. We check for the presence of photon pile--up using the SAS command {\sc epatplot}, and find no evidence for pile--up at any one of the three epochs. The background spectra are extracted from circular apertures of $\gtrsim$50\arcsec{} radii that are free from sources. These circular regions used to measure the background spectrum lie close to the source and on the same detector as the source.

When performing spectral analysis for \xmm{} data, we always jointly fit both the pn and the MOS spectra with the same fit function for the source spectra. To account for the instrument specific calibration differences, we use a constant component (\texttt{constant} in {\sc XSPEC}) multiplying the source models. This constant serves as a re--normalisation factor between different instruments. Specifically, we fix the constant to be 1 for the pn spectra, and let the constant for MOS ($C_{\rm MOS}$) free--to--vary in the fits for each epoch.

\subsection{\swf{} observations}

\swf{} performed 11 observations on the source J1231 between 2013 March and 2014 July. Following \citet{lin2017large}, we treat all \swf{} data as one epoch (Epoch S1). We combine the X--ray Telescope (\swf{}/XRT) data of all observations and extract the time--averaged, source$+$background and the background X--ray spectra of J1231 using the online XRT pipeline\footnote{https://www.swift.ac.uk/user$\_$objects/}, applying the default reduction criteria \citep{evans2009methods}.

\subsection{\chan{} observation}

\chan{} observed J1231 on 2016/02/10. We label the epoch as C1. We use CIAO (version 4.15) to perform the reduction of the data obtained by the Advanced CCD Imaging Spectrometer (ACIS) instrument onboard \chan{}. We employ the CIAO command {\sc chandra\_repro} for the data filtering, and {\sc specextract} for extraction of the spectrum. The source counts are extracted from a circular source region of 1.6\arcsec{} radius centred on the source (this of course also contains a small background contribution). This region corresponds to an encircled energy fraction of 95\% at 1~keV for an on--axis point source. The background spectrum is extracted from a circular region of $\sim$20\arcsec{} radius close to the source, on the same \chan/ACIS chip, and free from sources.

\section{Phenomenological characterisation of J1231 by blackbody models}

\begin{figure*}
   \centering
   \includegraphics[width=0.5\linewidth]{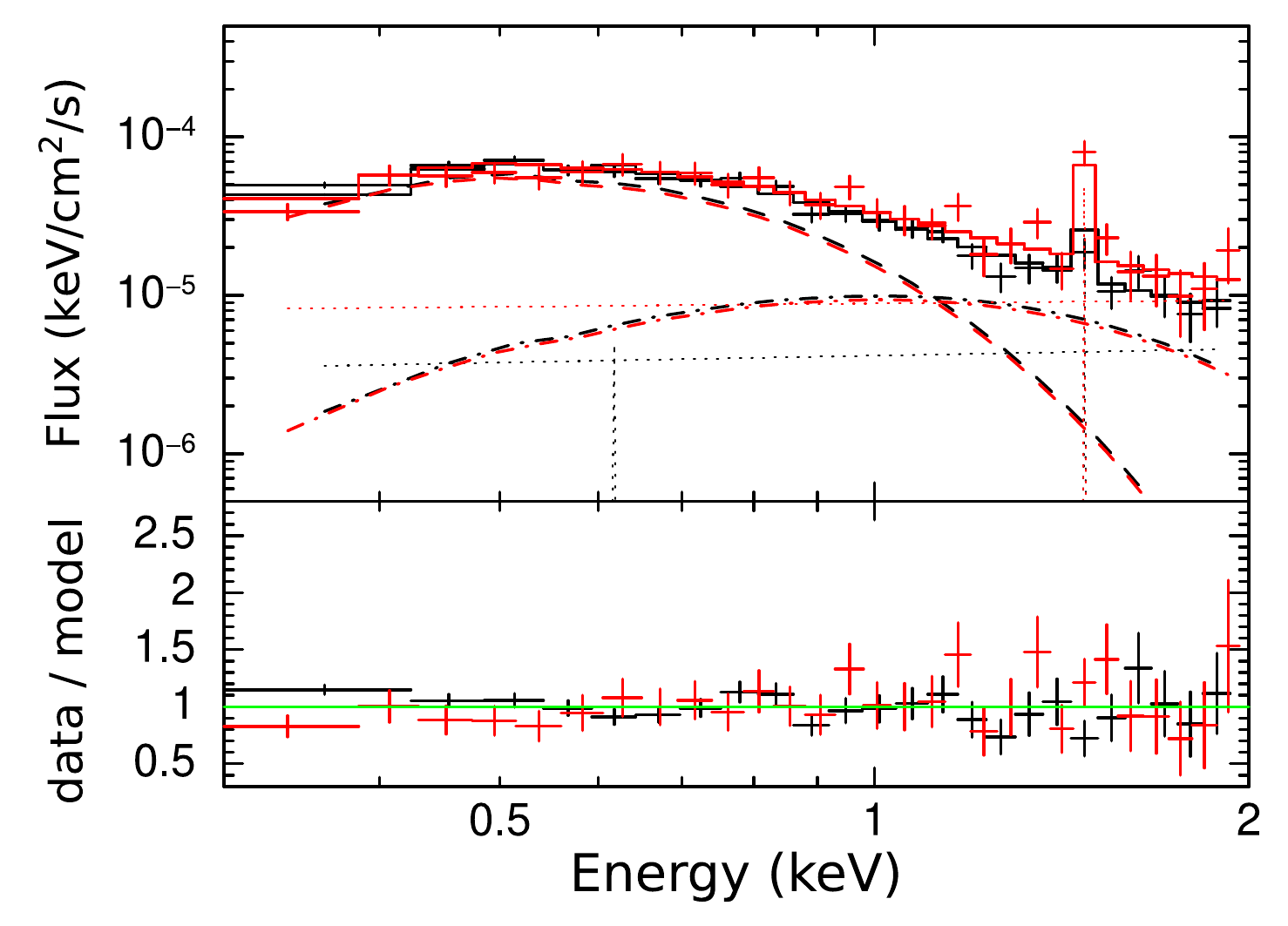}
   \caption{A phenomenological fit of two blackbodies (dashed and dot--dashed lines) to the \xmm{}/EPIC--pn (black) and \xmm/MOS (red) data from X2. The dotted lines represent the background spectra for each instrument. The bottom panel shows the ratio between the observed number of counts in each spectral bin (data; black and red points in the top panel) and the best--fit predicted number of counts in each spectral bin (model; solid lines in the top panel).}
   \label{fig:x2bb}
\end{figure*}

We characterise the spectra of J1231 at different epochs using simple blackbody models (\texttt{zbbody}; spectral models are referred to in the XSPEC syntax hereafter). For Epochs X1, S1, and C1, the source spectrum is consistent with a blackbody model (the total fit function \texttt{"constant*TBabs*zbbody"}), while for X2 and X3 a second, hotter blackbody component is required to achieve a good--fit to the data (for these epochs the total fit function becomes \texttt{"constant*TBabs*(zbbody+zbbody)"}). The two--blackbody--like spectra with $\dot m\sim\dot m_{\rm Edd}$ at Epochs X2 and X3 resemble several TDE spectra when accreting at high--Eddington or super--Eddington levels \citep[e.g.,][]{kara2018ultrafast,cao2023rapidly}. We summarise the best--fit parameters in Table~\ref{tb:2bb}. An example of the model fitted to the data at X2 is presented in Fig.~\ref{fig:x2bb}. The temperature of the primary blackbody component is $\sim$0.12~keV throughout the first four epochs before it drops to $0.07\pm0.02$~keV at the last epoch. For each epoch, we also test for the presence of intrinsic absorption (using the model \texttt{zTBabs}) and find that there is no evidence for significant intrinsic absorption (as also found by, e.g., \citealt{lin20133,lin2017large}). Therefore, we do not consider the intrinsic absorption in our analysis throughout the paper.

We also fit the time--resolved spectra produce in Section~\ref{sc:mdot} using \texttt{zbbody} models. Best--fit parameters are summarised in Table~\ref{tb:p&b}. The primary blackbody component at either X2 or X3, which dominates the 0.3--1.0~keV range, is consistent in temperature for the peak and valley composite spectra within the 1$\sigma$ uncertainty errors. However, it is not possible to constrain the second blackbody component of higher temperature in X2--v and X3--v due to the low number of spectral counts. Therefore, we cannot assess whether the variability introduces spectral shape changes above 1.0~keV.

\begin{table*}
\renewcommand{\arraystretch}{1.5}
\centering
\caption{Parameter constraints derived from fitting the spectrum at each epoch with one or two black bodies.}
\tiny
\begin{tabular}{cc|ccccc}
    \hline
    Model & Parameter & X1 & X2 & X3 & S1 & C1 \\
    \hline
    constant & $C_{\rm MOS}$ & $0.9\pm0.1$ & $0.94\pm0.05$ & $1.04\pm0.06$ & - & - \\
    \hline
    TBabs & $N_{\rm H}$ ($10^{20}$~cm$^{-2}$) & [2.6] & [2.6] & [2.6] & [2.6] & [2.6] \\
    \hline
    zbbody$_{\rm 1}$ & $kT_e$ (keV) & $0.116\pm0.004$ & $0.13\pm0.01$ & $0.13\pm0.01$ & $0.14\pm0.04$ & $0.07\pm0.02$ \\
    & Norm$_{\rm zbb1}$ ($10^{37}(1+z)^{-2}$erg$/$s$/$kpc$^2$)& $(1.51\pm0.09)\times10^{-6}$ & $(2.3\pm0.2)\times10^{-6}$ & $(2.0\pm0.1)\times10^{-6}$ & $(2\pm1)\times10^{-7}$ & $10_{-6}^{+22}\times10^{-7}$ \\
    \hline
    zbbody$_{\rm 2}$ & $kT_e$ (keV) & - & $0.28^{+0.10}_{-0.05}$ & $0.34_{-0.09}^{+0.19}$ & - & - \\
    & Norm$_{\rm zbb2}$ ($10^{37}(1+z)^{-2}$erg$/$s$/$kpc$^2$)& - & $(3_{-2}^{+3})\times10^{-7}$ & $(1.1_{-0.3}^{+0.7})\times10^{-7}$ & - & - \\
    \hline
    \multicolumn{2}{c|}{C-stat/d.o.f.} & 10.4/18 & 41.3/46 & 46.0/46 & 1.8/6 & 4.3/4\\
    \hline
\end{tabular}
\tablefoot{The fit function is \texttt{"constant*TBabs*(zbbody$_{\rm 1}$+zbbody$_{\rm 2}$)"}. Values held fixed during the fit are given in between square brackets. Parameter $C_{\rm MOS}$ is the re--normalisation factor between the EPIC/MOS and EPIC/pn on--board \xmm{}. Statistically, only X2 and X3 require the second, hotter black body to achieve a good fit, and therefore the fit function used to fit the other epochs do not include a second black body.}
\label{tb:2bb}
\end{table*}

\begin{table*}
\renewcommand{\arraystretch}{1.5}
\centering
\caption{Parameter constraints derived from fitting the spectrum at each epoch with one or two black bodies.}
\tiny
\begin{tabular}{cc|cc|cc}
    \hline
    Model & Parameter & X2--p & X2--v & X3--p & X3--v \\
    \hline
    constant & $C_{\rm MOS}$ & $0.94\pm0.07$ & $0.8\pm0.1$ & $0.97\pm0.08$ & $1.1\pm0.1$ \\
    \hline
    TBabs & $N_{\rm H}$ ($10^{20}$~cm$^{-2}$) & [2.6] & [2.6] & [2.6] & [2.6]  \\
    \hline
    zbbody$_{\rm 1}$ & $kT_e$ (keV) & $0.14\pm0.01$ & $0.14\pm0.01$ & $0.11\pm0.01$ & $0.13\pm0.01$\\
    & Norm$_{\rm zbb1}$ ($10^{37}(1+z)^{-2}$~erg$/$s$/$kpc$^2$) & $(3.6\pm0.3)\times10^{-6}$ & $(1.7\pm0.1)\times10^{-6}$ & $(2.7\pm0.3)\times10^{-6}$ & $(1.4\pm0.1)\times10^{-6}$\\
    \hline
    zbbody$_{\rm 2}$ & $kT_e$ (keV) & $0.3\pm0.2$ & - & $0.20\pm0.04$ & - \\
    & Norm$_{\rm zbb2}$ ($10^{37}(1+z)^{-2}$~erg$/$s$/$kpc$^2$) & $(4_{-1}^{+6})\times10^{-7}$ & - & $(6^{+6}_{-3})\times10^{-7}$ & - \\
    \hline
    \multicolumn{2}{c|}{C-stat/d.o.f.} & 52.0/43 & 14.3/17 & 35.3/42 & 25.1/18 \\
    \hline
\end{tabular}
\tablefoot{Same as Table~\ref{tb:2bb}, but here we fit the time--resolved spectra at the peaks (X2--p and X3--p) and the valleys (X2--v and X3--v). See Fig.~\ref{fig:lcs} for the production of those spectra. Due to the decrease of the source flux below the background level, we only consider the energy range of 0.3--1.0~keV for the X2--v and X3--v spectra in the analysis. It is not possible to detect the second, hotter black body in X2--v and X3--v. We find the best--fit temperature of the primary black body to be consistent with being the same in the peak and valley spectra.}
\label{tb:p&b}
\end{table*}

\section{Generating the time--resolved spectra for X2 and X3}

Here we list the time intervals selected for stacking the spectra of the peaks (X2--p and X3--p) and the valleys (X2--v and X3--v), as presented in Fig.~\ref{fig:lcs}. The intervals are given in offsets in seconds from the \xmm{} Mission Reference Time (MRT, 1997-12-31T23:58:56.816 UTC), minus 2.50860675e+08~s. Both the pn and the MOS data use the same time intervals to produce the stacked spectra.

For X2--p spectrum: from 0 to 5500; from 15300 to 19700; from 28500 to 31500; from 39500 to 44500; from 56000 to 60000.

For X2--v spectrum: from 7000 to 13300; from 21500 to 27000; from 33500 to 36800; from 48000 to 52000.

For X3--p spectrum: from 328151 to 333151; from 342651 to 345151; from 356651 to 360651; from 369151 to 373151; from 382651 to 387151; from 397151 to 403151.

For X3--v spectrum: from 335151 to 340151; from 349151 to 354151; from 361651 to 366651; from 376651 to 380651; from 389651 to 394651; from 406151 to 410151.

\section{Tables of fit parameter constraints}

\begin{table*}
\renewcommand{\arraystretch}{1.5}
\centering
\caption{Parameter constraints for the joint fit of the spectra at all the epochs.}
\tiny
\begin{tabular}{cc|ccccc}
    \hline
    Model & Parameter & X1 & X2 & X3 & S1 & C1 \\
    \hline
    constant & $C_{\rm MOS}$ & $0.9\pm0.1$ & $0.93\pm0.05$ & $1.04\pm0.05$ & - & - \\
    & $C_{\rm XRT}$ or $C_{\rm ACIS}$ & - & - & - & [1] & [1] \\
    \hline
    TBabs & $N_{\rm H}$ ($10^{20}$~cm$^{-2}$) & \multicolumn{5}{c}{[2.6]} \\
    \hline
    slimdz & $\dot m$ ($\dot m_{\rm Edd}$) & $0.43_{-0.02}^{+0.20}$ & $0.93_{-0.03}^{+0.54}$ & $0.65_{-0.02}^{+0.33}$ & $0.13\pm0.04$ & $0.10\pm0.02$\\
    & $\theta$ ($^{\circ}$)& \multicolumn{5}{c}{$<30$}\\
    & $M_{\bullet}$ ($M_{\odot}$)& \multicolumn{5}{c}{$(5.7_{-0.8}^{+0.5})\times10^4$}\\
    & $a_{\bullet}$ & \multicolumn{5}{c}{$>0.96$}\\
    \hline
    \multicolumn{2}{c|}{C-stat/d.o.f.} & \multicolumn{5}{c}{136.9/124}\\
    \hline
\end{tabular}
\tablefoot{The fit function is \texttt{"constant*TBabs*slimdz"}. Values held fixed during the fit are given in square brackets. During the fit, the $M_{\bullet}$, $a_{\bullet}$, and $\theta$ are free--to--vary but forced to be the same at each epoch. Due to low number of photons detected at S1 and C1, it is not possible to simultaneously constrain the re--normalisation factor ($C_{\rm XRT}$ and $C_{\rm ACIS}$) and $\dot m$. Thus we fix $C_{\rm XRT}$ and $C_{\rm ACIS}$ to unity, assuming different instruments are well cross--calibrated. The expected C-stat is calculated to be $C_e\approx$135, with a root--mean--square (RMS) of $\approx$16. We find that the X--ray spectra of J1231 at all epochs can be well explained by a slim disc model varying only the mass accretion rate $\dot m$.}
\label{tb:jointfit}
\end{table*}

\begin{table*}
\renewcommand{\arraystretch}{1.5}
\centering
\caption{Parameter constraints for the joint fit of the spectra at all the epochs.}
\tiny
\begin{tabular}{cc|ccccc}
    \hline
    Model & Parameter & X1 & X2 & X3 & S1 & C1 \\
    \hline
    constant & $C_{\rm MOS}$ & $0.9\pm0.1$ & $0.93\pm0.05$ & $1.03\pm0.05$ & - & - \\
    & $C_{\rm XRT}$ or $C_{\rm ACIS}$ & - & - & - & [1] & [1] \\
    \hline
    TBabs & $N_{\rm H}$ ($10^{20}$~cm$^{-2}$) & \multicolumn{5}{c}{[2.6]} \\
    \hline
    slimdz & $\dot m$ ($\dot m_{\rm Edd}$) & $11_{-4}^{+8}$ & =X1 & =X1 & $0.13_{-0.03}^{+0.08}$ & $0.10\pm0.03$\\
    & $\theta$ ($^{\circ}$)& $76\pm3$ & $71\pm3$ & $73\pm3$ & $<65$ & =S1\\
    & $M_{\bullet}$ ($M_{\odot}$)& \multicolumn{5}{c}{$(5\pm1)\times10^4$}\\
    & $a_{\bullet}$ & \multicolumn{5}{c}{$>0.97$}\\
    \hline
    \multicolumn{2}{c|}{C-stat/d.o.f.} & \multicolumn{5}{c}{119.0/123}\\
    \hline
\end{tabular}
\tablefoot{Same as Table~\ref{tb:jointfit}, but here we find that the X--ray spectra of J1231 at X1/X2/X3 epochs can also be well explained by a slim disc model varying only the inclination $\theta$. Although the marginal constraints of $\theta$ at X1/X2/X3 overlaps, for any given set of \{$\dot m$, $M_{\bullet}$, $a_{\bullet}$\}, they do not overlap. During the fit, we assume at S1 and C1, decades after the first detection, any possible disc precession is terminated, and so the disc inclination is not varying. Based on the calculation of $C_e$ and its RMS, both fits (Table~\ref{tb:jointfit} and this table) should be considered good descriptions of the data.}
\label{tb:jointfit-theta}
\end{table*}

\begin{table*}
\renewcommand{\arraystretch}{1.5}
\centering
\caption{Parameter constraints from fitting the time--resolved spectra at X2 and X3 with the \texttt{slimdz} model.}
\tiny
\begin{tabular}{cc|cc|cc}
    \hline
    Model & Parameter & X2--p & X2--v & X3--p & X3--v \\
    \hline
    constant & $C_{\rm MOS}$ & $0.94\pm0.07$ & $0.8\pm0.1$ & $0.97\pm0.08$ & $1.1\pm0.1$ \\
    \hline
    TBabs & $N_{\rm H}$ ($10^{20}$~cm$^{-2}$) & \multicolumn{2}{c|}{[2.6]} & \multicolumn{2}{c}{[2.6]} \\
    \hline
    slimdz & $\dot m$ ($\dot m_{\rm Edd}$) & $1.74\pm0.08$ & $0.64\pm0.03$ & $1.05\pm0.04$ & $0.55\pm0.02$\\
    & $\theta$ ($^{\circ}$)& \multicolumn{2}{c|}{[10]} & \multicolumn{2}{c}{[10]}\\
    & $M_{\bullet}$ ($M_{\odot}$)& \multicolumn{2}{c|}{[$5.7\times10^4$]} & \multicolumn{2}{c}{[$5.7\times10^4$]}\\
    & $a_{\bullet}$ & \multicolumn{2}{c|}{[0.99]} & \multicolumn{2}{c}{[0.99]}\\
    \hline
    \multicolumn{2}{c|}{C-stat/d.o.f.} & \multicolumn{2}{c|}{74.9/64} & \multicolumn{2}{c}{74.7/64}\\
    \hline
\end{tabular}
\tablefoot{The total fit function is \texttt{"constant*TBabs*slimdz"}. Values held fixed during the fit are given in square brackets. Assuming the best--fit BH parameters ($M_{\bullet}$, $a_{\bullet}$, and $\theta$) derived from Table~\ref{tb:jointfit}, we test the disc explanation of the quasi--periodic variation observed at X2 and X3. At each epoch of X2 and X3, we jointly fit the average spectrum of the peaks (X2--p or X3--p) and that of the valleys (X2--v or X3--v). Here $C_e\approx68$ with an RMS of $\approx12$ for each epoch. We find the difference between the source spectra at peaks and valleys can be explained by a varying $\dot m$.}
\label{tb:mdot}
\end{table*}

\begin{table*}
\renewcommand{\arraystretch}{1.5}
\centering
\caption{Parameter constraints from fitting the time--resolved spectra at X2 and X3 with the \texttt{slimdz} model.}
\tiny
\begin{tabular}{cc|cc|cc}
    \hline
    Model & Parameter & X2--p & X2--v & X3--p & X3--v \\
    \hline
    constant & $C_{\rm MOS}$ & $0.96\pm0.07$ & $0.8\pm0.1$ & $0.97\pm0.08$ & $1.1\pm0.1$ \\
    \hline
    TBabs & $N_{\rm H}$ ($10^{20}$~cm$^{-2}$) & \multicolumn{2}{c|}{[2.6]} & \multicolumn{2}{c}{[2.6]} \\
    \hline
    slimdz & $\dot m$ ($\dot m_{\rm Edd}$) & $17\pm2$ & $2.5\pm0.2$ & $5.7\pm0.4$ & $2.1\pm0.1$\\
    & $\theta$ ($^{\circ}$)& \multicolumn{2}{c|}{[10]} & \multicolumn{2}{c}{[10]}\\
    & $M_{\bullet}$ ($M_{\odot}$)& \multicolumn{2}{c|}{[$3\times10^4$]} & \multicolumn{2}{c}{[$3\times10^4$]}\\
    & $a_{\bullet}$ & \multicolumn{2}{c|}{[0.6]} & \multicolumn{2}{c}{[0.6]}\\
    \hline
    \multicolumn{2}{c|}{C-stat/d.o.f.} & \multicolumn{2}{c|}{85.6/64} & \multicolumn{2}{c}{71.4/64}\\
    \hline
\end{tabular}
\tablefoot{Same as Table~\ref{tb:mdot}, but here we assume the case of the smallest spin value based on results in Fig.~\ref{fig:contour} ($M_{\bullet}=3\times10^4$~$M_{\odot}$ and $a_{\bullet}=0.6$). The total C-stat of X2 and X3 is larger than that shown in Table~\ref{tb:mdot}, due to the choice of \{$M_{\bullet}$, $a_{\bullet}$\} is 2$\sigma$ away from the best--fit case.}
\label{tb:smallspin}
\end{table*}

\begin{table*}
\renewcommand{\arraystretch}{1.5}
\centering
\caption{Parameter constraints from fitting the time--resolved spectra at X2 and X3 with the \texttt{slimdz} model.}
\tiny
\begin{tabular}{cc|cc|cc}
    \hline
    Model & Parameter & X2--p & X2--v & X3--p & X3--v \\
    \hline
    constant & $C_{\rm MOS}$ & $0.94\pm0.07$ & $0.8\pm0.1$ & $0.98\pm0.08$ & $1.1\pm0.1$ \\
    \hline
    TBabs & $N_{\rm H}$ ($10^{20}$~cm$^{-2}$) & \multicolumn{2}{c|}{[2.6]} & \multicolumn{2}{c}{[2.6]} \\
    \hline
    slimdz & $\dot m$ ($\dot m_{\rm Edd}$) & $10_{-8}^{+11}$ & $6\pm2$ & $24\pm5$ & $6\pm1$\\
    & $\theta$ ($^{\circ}$)& $61_{-39}^{+3}$ & $77\pm1$ & $68\pm1$ & $78\pm1$\\
    & $M_{\bullet}$ ($M_{\odot}$)& \multicolumn{2}{c|}{[$5\times10^4$]} & \multicolumn{2}{c}{[$5\times10^4$]}\\
    & $a_{\bullet}$ & \multicolumn{2}{c|}{[0.99]} & \multicolumn{2}{c}{[0.99]}\\
    \hline
    \multicolumn{2}{c|}{C-stat/d.o.f.} & \multicolumn{2}{c|}{68.5/64} & \multicolumn{2}{c}{63.8/64}\\
    \hline
\end{tabular}
\tablefoot{Same as Table~\ref{tb:mdot}, but here we assume the best--fit BH parameters ($M_{\bullet}=5\times10^4$, $a_{\bullet}=0.99$) derived from the early $\theta$--varying scenario (Table~\ref{tb:jointfit-theta}). Good fits to both X2 and X3 spectra can only be achieved by letting both $\dot m$ and $\theta$ free to vary between the peaks and the valleys.}
\label{tb:highincl}
\end{table*}

%%%%%%%%%%%%%%%%%%%%%%%%%%%%%%%%%%%%%%%%%%%%%%%%%%
\end{appendix}
\newpage
% Don't change these lines
\label{lastpage}
\end{document}